\documentclass[11pt]{article}

\usepackage[utf8x]{inputenc}

\usepackage[a4paper]{geometry}
\usepackage{doi}
\usepackage{hyperref}
\usepackage[sort&compress,square,comma,numbers]{natbib}

\usepackage{lipsum}
\usepackage{amsmath}
\usepackage{amsfonts}
\usepackage{graphicx}
\usepackage{epstopdf}
\usepackage{algorithmic}
\ifpdf
  \DeclareGraphicsExtensions{.eps,.pdf,.png,.jpg}
\else
  \DeclareGraphicsExtensions{.eps}
\fi

\renewcommand{\S}{$\S$}
\newcommand{\T}{\mathcal{T}}

\newcommand{\M}{\ensuremath{\mathbb{M}}}

\title{ 
The dynamics of a collapsing polyelectrolyte gel
}

\author{Giulia~L.~Celora%
\thanks{Mathematical Institute, Woodstock Road, University of Oxford, Oxford, OX2 6GG, UK 
}
\and
Matthew G.~Hennessy\footnotemark[1]
\and
Andreas~M\"unch\footnotemark[1]
\and
Barbara~Wagner\thanks{Weierstrass Institute, Mohrenstrasse 39, 10117 Berlin, Germany
}
\and
Sarah~L.~Waters\footnotemark[1]
}

\usepackage{amsopn}

\ifpdf
\hypersetup{
  pdftitle={
  The dynamics of collapse of a polyelectrolyte gel},
  pdfauthor={G. Celora, M. Hennessy, A. M\"unch, B. Wagner, S. Waters}
}
\fi

\usepackage{our_commands}
\usepackage{multirow}
\usepackage{booktabs}
\usepackage[labelfont=bf]{caption}
\usepackage{subcaption}
\usepackage[makeroom]{cancel}
\usepackage{afterpage}
\usepackage{todonotes}
\newcommand{\ionbp}{\ensuremath{\phi_{0^+}}}

\renewcommand{\d}{\ensuremath{\text{d}}}

\allowdisplaybreaks

\begin{document}

\maketitle
\begin{abstract}
We analyse the dynamics of different routes to collapse of a constrained polyelectrolyte gel in contact with an ionic bath. The evolution of the gel is described by a model that incorporates non-linear elasticity, Stefan-Maxwell diffusion and interfacial gradient free energy to account for phase separation of the gel. 
A bifurcation analysis of the homogeneous equilibrium states reveals three solution branches at low ion concentrations in the bath, giving way to only one above a critical ion concentration.
We present numerical solutions that capture both the spatial heterogeneity and the multiple time-scales involved in the process of collapse. These solutions are complemented by two analytical studies. Firstly, a phase-plane analysis that reveals the existence of a depletion front for the transition from the highly swollen to the new collapsed equilibrium state. This depletion front is initiated after the fast ionic diffusion has set the initial condition for this time regime. Secondly, we perform a linear stability analysis about the homogeneous states that show 
that for a range of ion concentrations in the bath, spinodal decomposition of the swollen state gives rise to localized solvent-rich(poor) and, due to the electro-neutrality condition, ion-poor(rich) phases that coarsen on the route to collapse. This dynamics of a collapsing polyelectrolyte gel has not been described before. 
\end{abstract}
\section{Introduction}
Ever since the seminal papers by Tanaka et al. \cite{Tanaka1978} and Dusek \cite{Dusek1968}, research on swelling and collapse of polyelectrolyte gels has been very intensive, both theoretically and experimentally \cite{dobrynin_theory_2008,Dimitriyev_2019,Horkay2001,mccoy_dynamic_2010,mussel_experimental_2019}. These systems, combining elements of electrochemistry and condensed matter physics, display intriguing and subtle properties motivating both experimental and theoretical studies to understand their rich behaviour. These gels also have a wealth of technological applications, and a better understanding of polyelectrolyte gels serves as a basis for developing smart, responsive materials and sensors \cite{Buenger2012,Chaterji2007,Hong2012,Stuart2010}, for example. In particular, research in this field is driven by applications in medicine \cite{Hong2012,Kwon2006}, e.g. for drug delivery and tissue engineering. Additionally, polyelectrolyte gels are used as a model system for many types of biological tissues \cite{Lutolf2005,Ning2018} to gain fundamental insight into diverse phenomena in biology. Polyelectrolytes also serve as a model for bio-macromolecules such as DNA and RNA \cite{estevez-torres_dna_2011,roshal_viral_shell_2019,zandi_virus_2020}. 

In its simplest form a polyelectrolyte gel is a network of covalently cross-linked polyelectrolyte macromolecules, that is, of polymer chains carrying fixed charges of the same sign, immersed in a solvent. If placed in an ionic bath, the gel will approach a new equilibrium state driven by osmotic effects and will swell or shrink \cite{Horkay2001}. 
This process depends on factors such as the concentration and valency of the salt in the solvent, the (nonlinear) elasticity of the gel, the concentration of fixed charges and the number of ionizable groups of the polyelectrolyte macromolecule, as well external fields such an applied electric field or the temperature \cite{Kokufuta2002}. 
The change in volume does not always proceed continuously \cite{Budtova1998,mussel_experimental_2019, Yu2017}. The 
discontinuous volume phase transition typically considers a gel that is divided into co-existing subdomains that are in thermodynamic equilibrium with jump conditions imposed at their interfaces \cite{Doi2009, Cai2011}.

Unlike non-ionic hydrogels \cite{Hennessy2020,
Sato1988, Tomari1995}, subtle changes in the environment surrounding the gel
such as increasing the ion concetration
can have a dramatic effect and result in discontinuous phase
transitions connected with super-collapse
\cite{Khokhlov_PolyelecCollape_1996,hua_theory_2012} and 
re-entrant swelling \cite{sing_reentrant_swelling_2013}. 
A deeper understanding of these phenomena, in particular when comparing to
experiments, is obtained with a model that resolves the pattern forming
instabilities of the gel and the transient dynamics between equilibrium states
over a large range of temporal and spatial scales.  This will then shed light
on the pattern formation processes leading to collapse
\cite{matsuo_patterns_1992,zubarev_self-similar_2004}. 

The governing equations of such a model are given in a companion paper
\cite{Celora_modelling_2020}, where we use non-equilibrium thermodynamics to
systematically derive a phase-field model of a polyelectrolyte gel.  That model
accounts for the free energy of the internal interfaces which form upon phase
separation, as well as for finite elasticity, together with multi-component transport models via Stefan-Maxwell diffusion. We also derive a thermodynamically consistent model for the ionic bath surrounding the gel. The electro-neutral limit of
the full 3D model has been derived via matched asymptotic expansions  \cite{Celora_debye_2020} from which we obtain the jump conditions that need to be imposed at the gel-bath interface. First results from numerical simulations for a 1D constrained gel have been presented in \cite{Celora_modelling_2020}, where the possibility of  spinodal decomposition is raised for the case of a \emph{swelling} gel.

The main goal of this study is to investigate the evolution of
collapse via a volume phase transition in a constrained gel by a combination of mathematical techniques. 
First, we consider the nonlinear equations that determine the homogeneous
equilibrium states and deduce the bifurcation that leads to the collapse as the
salt concentration in the bath is raised. The numerical solution then
reveals the fast initial transients that change the charge distribution in the
gel, followed by the appearance of a (solvent) depletion front. This in turn we study via a
phase-plane analysis. We systematically investigate the stability  
of the homogeneous states. This reveals a second route to
collapse, where the gel first undergoes spinodal decomposition before the
depletion front moves through the heterogeneous state.

The paper is structured as follows. In Section \ref{sec_formulation} we introduce the electroneutral version of the model for a polyelectrolyte gel, previously derived in  \cite{Celora_modelling_2020}. In Section \ref{sec:1Dmodel} we specialise the model to the case of a one-dimensional constrained gel in contact with an ionic 
bath, where the thin electric double layer between gel and bath, is replaced by corresponding boundary conditions \cite{Celora_debye_2020}.
In Section \ref{sec_eq}, we determine the bifurcation diagram for the homogeneous equilibrium states which are controlled by the salt concentration in the bath,
and give an overview of various routes to collapse. We carry out numerical simulations that demonstrate the fast dynamics of the free ions that sets the stage for collapse via a moving depletion front. In Section~\ref{sec_collapse} we show the existence of this moving front via a phase plane analysis. Section~\ref{sec_stability} investgates different parameter regimes that show spinodal decomposition. We use linear stability analysis that identify paramter regimes for spinodal decomposition into localized solvent-rich(poor) phases that eventually coarsen on their path to collapse. In Section \ref{sec:conclusions} we draw our conclusions and give an outlook on further research directions.

\section{Governing equations for a polyelectrolyte gel}\label{sec_formulation}
We consider the problem of a polyelectrolyte gel swelling in a solution (the bath) containing a binary salt. The governing equations are systematically derived in
\cite{Celora_modelling_2020}, which extends previous models by
Drozdov et al.~\cite{Drozdov2016b} and Hong et al.~\cite{Hong2010}. In particular, we include the gradient energy to account for the dynamics of collapse and phase separation.
As standard in the literature, we consider the electroneutral formulation of the model. In Hennessy et al. \cite{Celora_debye_2020} we give a detailed derivation of this limit via singular perturbation analysis, where we assume that Debye length and thus the thickness of the double-layer at the free interface with the gel with the bath is small compared to the size of the gel. 

The governing equation are presented for the full 3D problem in terms of the Eulerian coordinates associated with the current (deformed) configuration.
The gel is described as a mixture of three phases: solvent ($s$), free ions ($+$ and $-$) and the charged polymer network ($n$). We assume that the fixed charges (with valence $z_f$) on the polymer network are evenly distributed on the network and account for a fixed fraction $\alpha_f<1$ of the network volume. In this work we consider $z_f$ to be positive so that the positive ionic species  in the solution (with valence $z_+$) will be denoted as co-ions ($+$) while the negative ones (with valence $z_-$) will be the counter-ions ($-$). As standard in the polyelectrolyte gel literature, we will assume in what follows that all components have the same molecular volume $\nu$. 

Let us denote by $t$ time and $\vec x=(x,y,z)$ the Eulerian coordinates, while $\vec{X}=\vec{X}(\vec{x},t)$ are the corresponding Lagrangian coordinates. Then the gel kinematics are described by the deformation gradient tensor $\tens F=(\partial \vec X/\partial \vec x)^{-1}$, with its determinant $J=\det \tens F$ representing the volume expansion of the gel compared to the dry reference
state, where $J=1$. The velocity of the polymer network
$\vec v_n$ can be expressed as
\begin{align}
\vec v_n= -\tens{F}\frac{\partial \vec X}{\partial t}.
\end{align}
The composition of the gel is described in terms of volume fractions $\phi_i=\phi_i(\vec{x},t)$ with $i \in\left\{s,+,-,n\right\}$. Assuming there are no voids in the gel, the volume fractions satisfy 
\begin{subequations}\label{model3D1}
\begin{align}
1= \phi_n+\phi_s+\phi_++\phi_-.\label{eq:novoid}
\end{align}
Moreover, all of the phases are assumed to be incompressible; consequently,
$J$ is related to the network
volume fraction $\phi_n$ by $ J=\phi_n^{-1}$.
The volume fraction occupied by fixed charges is
$\phi_f=\alpha_f \phi_n$.
Since we assume the gel is electrically neutral at each location in space, its net charge must be zero. This introduces the additional constraint
\begin{align}
z_f\phi_f +z_+\phi_++z_-\phi_-&=0. \, \label{eq:electroneutral}
\end{align}
Conservation of mass and momentum are given by
\begin{align}
\partial_t \phi_n + \nabla \cdot(\phi_n \vec{v}_n) &=0, \label{eq:mass_cn}\\
\partial_t \phi_m + \nabla \cdot(\phi_m \vec{v}_n) &=-\nabla \cdot\vec{j}_m, \label{eq:mass_cons}\\
  \nabla\cdot\tens{T} &= 0,\label{eq:stressbalance}
\end{align}
for $m \in \M = \left\{s,+,-\right\}$. Here, $\tens T$ denotes the Cauchy stress tensor (to be specified below). The diffusive fluxes $\vec j_m=\vec j_m(\vec x,t)$  are defined as the volumetric flux of mobile species relative to the network, i.e. $\vec{j}_m=\phi_m(\vec v_m -\vec v_n)$ with $\vec v_m$ being the velocity of the $m$-th mobile component of the mixture.
By differentiating~(\ref{eq:novoid}) and~(\ref{eq:electroneutral}) with respect to time and using~(\ref{eq:mass_cn})--(\ref{eq:mass_cons}) we obtain
\begin{align}
    \nabla \cdot\left(\vec v_n+\sum_{m\in \M} \vec j_m\right)=0,\label{eq:condvandfluxes}\\
    \nabla \cdot\left(z_+\vec{j}_++z_-\vec j_-\right)=0\label{eq:condfluxes},
\end{align}
\end{subequations}
which are used to replace \eqref{eq:mass_cn} and \eqref{eq:mass_cons} for the
counter-ions fraction $\phi_{-}$. 

To complete the model, constitutive relations are required. The fluxes are described via Stefan-Maxwell diffusion to account for the relative friction between ions and solvent,
\begin{subequations}\label{constlaws}
\begin{align}
\vec{j}_s &= -\phi_s K  \left(\nabla \mu_s + \frac{\phi_+}{\phi_s} \nabla \mu_+
+ \frac{\phi_-}{\phi_s} \nabla \mu_-
\right),\\
\vec j_\pm&=-\frac{\D_\pm \phi_\pm}{k_BT}\nabla \mu_\pm + \frac{ \phi_\pm}{\phi_s}\vec{j}_s.
\end{align}
where $\mu_s$, $\mu_+$ and $\mu_-$ denote
the (electro)chemical potentials of the mobile species, $T$ is the temperature, $k_B$ is the Boltzmann constant, $K=\D_s/(k_BT)$ is the solvent permeability and $\D_m$ are the diffusivities of the mobile ions, where we consider $\D_+=\D_->\D_s$ in line with the characteristic values reported (see supplementary material Table~\ref{tab:physparms}).   
Furthermore, we have the following expressions for the three chemical potentials
\begin{align}
&\mu_s = \mu_s^0 + \nu p - \frac{\gamma}{\nu}\nabla^2\phi_s + k_BT\left[\ln (\phi_s)  + \frac{\chi(1-\phi_s)}{J} + \frac{1}{J}
\right], \label{eqn:mu_s}\\
&\mu_\pm = \mu^0_\pm + \nu p + z_\pm e\Phi + k_BT \left[\ln (\phi_\pm)-\frac{\chi \phi_s}{J} 
+ \frac{1}{J} \right],
\end{align}
where $\mu^0_{m}$ are reference chemical potentials, $\Phi$ is the electric potential, $p$ is the pressure, $e$ is the elementary charge,
$\chi$ is the temperature-dependent
Flory interaction parameter and
$\gamma$ is the interfacial stiffness parameter. The latter is associated with the free energy cost of internal interfaces separating regions of low and high solvent concentration.
For the stresses in the gel, we consider three contributions
\begin{align}
&\tens{T} = -p \tens{I} +  \tens{T}_K+ \tens{T}_E.\label{Tconst}
\end{align}
The first represents an isotropic stress from the fluid pressure
with $\tens{I}$ denoting the
identity tensor, while $\tens{T}_K$ represents 
the Korteweg stress generated at internal interfaces 
(i.e., gradients of the solvent concentration)
\begin{align}
\tens{T}_K &=\frac{\gamma}{\nu^2}\left[\left(\frac{|\nabla \phi_s|^2}{2}+\phi_s\nabla^2\phi_s\right)\tens{I} - \nabla \phi_s \otimes \nabla \phi_s\right], 
\intertext{and $\tens T_E$ is the elastic stress from the 
response of a neo-Hookean polymer network}
\tens{T}_E&=\frac{G\left(\tens{B}-\tens{I}\right)}{J}.
\end{align}
where $\tens{B}=\tens F \tens F^T$ is the
left Cauchy-Green tensor and $G$ is the shear modulus. 
\end{subequations}
To close the system, we further need boundary and initial conditions which we derive next for the problem of a constrained gel, which is the focus of this study. 

\section{Specialisation to constrained swelling and collapse}
\label{sec:1Dmodel} 

\begin{figure}[tb]
		\centering
		\includegraphics[width=0.9\textwidth]{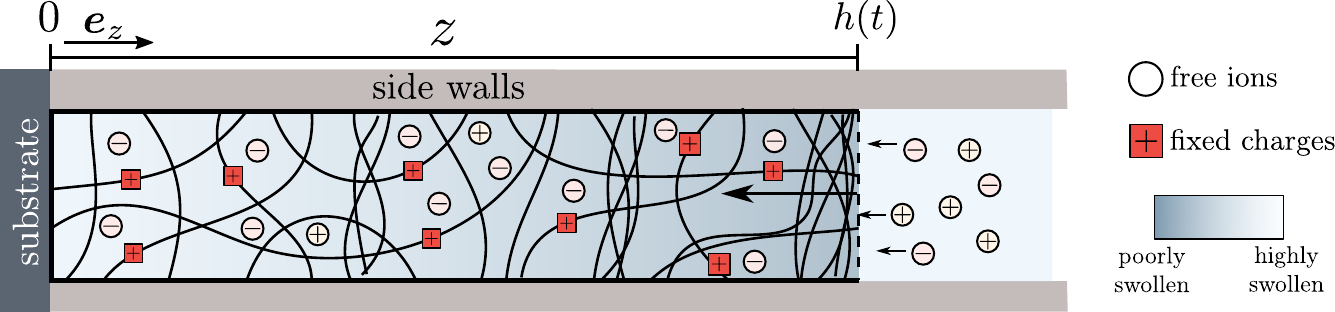}
		\caption{Schematic of a laterally confined polyelectrolyte gel that collapse along the $z$-axis only. The free interface with the bath is located at $z=h(t)$. As solvent is expelled by the gel the free interface moves towards the left and salt ions are  absorbed/desorbed by the gel so as to maintain electro-neutrality of the gel.}
		\label{schematic}
\end{figure}

We consider the case of a constrained gel which undergoes uni-axial deformation (in the $z$ direction) due to the uptake or release of a monovalent ($z_\pm=\pm 1$) salt solution (see Figure \ref{schematic}). This is analogous to the scenario considered in~\cite{Doi2009,Hennessy2020} for the study of constrained swelling and deswelling of neutral hydrogels.

The gel is assumed to be attached to a substrate at $z=0$ while the interface at $z=h(t)$ is free to move along frictionless side walls that do not influence the bulk behaviour (see \cite{Hennessy2020} for more details). 
The problem can be reduced to a 1D Cartesian geometry, where the deformation gradient tensor $\tens{F}$  and the stress tensor $\tens{T}$ have the form
\begin{equation}
\tens{F}=\mbox{diag}(1,1,J(z,t)), \quad \tens{T}=\mbox{diag}(T_\ell(z,t),T_\ell(z,t),T_{zz}(z,t)),
\end{equation}
and the diffusive fluxes and network velocity are given by
$\vec{j}_m=j_m(z,t) \vec{e}_z$ and $\vec{v}_n = v_n(z,t) \vec{e}_z$ respectively, with
$\vec e_z$ representing the unit vector in the $z$ direction. Moreover,
all variables in (\ref{model3D1})-(\ref{constlaws}) are assumed to only depend on $z$ and $t$.
At $z=0$, we assume that the gel is bounded by a solid, {insulated}, impermeable substrate;
hence,
\begin{subequations}\label{bottombc}
\begin{align}
v_n(0,t)=0, \quad j_s(0,t)=0, \quad j_\pm(0,t)=0.
\end{align}
Given the fourth-order derivative in the solvent fraction that ultimately arises from
the interfacial term in the solvent chemical potential \eqref{eqn:mu_s},
we need an additional condition on $\phi_s$. We assume the solvent is neutral to
substrate, resulting in \cite{puri_surface-directed_2002}
\begin{align}
\quad \partial_z \phi_s(0,t)=0.
\end{align}
\end{subequations}
In most scenarios considered in this paper, we assume that the free surface of the gel at $z=h(t)$ is in contact with an ionic and electro-neutral bath, which behaves like an infinite reservoir of ions. For this scenario, we adapt the boundary conditions derived in \cite{Celora_debye_2020} starting from the full non-electroneutral model via singular perturbation analysis to account for the formation of a small layer, known as the electric double layer, near the free interface where electro-neutrality breaks down. Away from the interface with the gel, the volume fraction of ions in the bath is controlled experimentally and set to the value $\phi_0$. In our one-dimensional situation, and assuming the electric potential is zero in the bath, we have
\begin{subequations}\label{bbc}
\begin{align} 
T_{zz}  (h(t),t)&=0,\label{cond:nostress}\\
\partial_z \phi_s (h(t),t)&=0, \label{cond:var}\\
\mu_\pm (h(t),t)&=\mu^0_\pm+\ln(\phi_0),\label{mupmcont2}\\
\mu_s (h(t),t)&=\mu^0_s+\ln(1-2\phi_0).\label{muscont}
\end{align}
Note that for a monovalent salt solution, we can rewrite the conditions~(\ref{mupmcont2}) as:
\begin{align}
    \bar{\mu} (h(t),t)&=\mu^0_++\mu^0_-+2\ln(\phi_0)\label{mupmcont},\\
    \Phi(h(t),t)&=\left.\frac{k_BT}{e}\ln\left( \frac{\phi_-}{\phi_+}\right)\right|_{z=h(t)}\label{bc:Phi}
\end{align}
where~(\ref{mupmcont}) and~(\ref{bc:Phi}) are obtained respectively summing and subtracting~(\ref{mupmcont2}). Here $\bar{\mu}$ is defined as $\bar{\mu}=\mu_++\mu_-$ and it is referred to as the chemical potential of the free ions. To close the system, we further specify a kinematic condition for the boundary $z=h(t)$, which here moves with the gel velocity such that
\begin{align}
	\frac{d h}{dt}=v_n(h(t),t)
\end{align}
\end{subequations}

Having specified the geometry and boundary conditions for the problem, we non-dimensionalise the system as follows:
\begin{equation}
\begin{aligned}
\hat\mu_m = \frac{\mu_m-\mu^0_m}{k_BT}, &\quad \hat \Phi = \frac{\Phi e}{k_B T}, \quad
\hat{\tens{T}}=\frac{\tens{T}}{G}, \qquad \hat{z} =\frac{z}{L}, \\
 \qquad \hat t=\frac{t}{\tau},&\quad \hat p= \frac{p}{G}, \qquad \hat{j}_m=\frac{\nu L}{\D_s}j_m, \qquad \tau=\frac{L^2}{\D_s},
 \end{aligned}
\end{equation}
where $m\in\mathbb{M}$ and $L$ is the characteristic size of the gel. 
The system possesses a second natural length scale,
$L_{\text{int}}=\sqrt{\gamma/\left(\nu k_BT\right)}$,
which characterises the thickness of the internal interfaces that can occur via phase separation of the gel into highly and poorly swollen regions. The ratio of these
 two length scales gives rise to the non-dimensional parameter 
$\omega= L_{\text{int}}/L$.
Further non-dimensional material parameters are 
\begin{equation}
\mathcal{G}=\frac{\nu G}{k_BT}, \quad  \hat{\mathcal{D}} =\frac{\D}{\D_s}.
\end{equation}
Note that $\G$ can be related to the number density of polymer chains $N_p$ in the dried network via $\G=N_p\nu$, which helps in the estimation of its value (see Table~\ref{tab:physparms}).
We introduce the scalings into the 1D model and then 
drop the hat notation, so that the non-dimensional governing equations read
\begin{subequations}
\begin{align}
\partial_t \phi_s + \partial_z(\phi_s v_n)&= - \partial_z j_s,\label{fluxs}\\
\partial_t \phi_+ + \partial_z(\phi_+ v_n)&= -\partial_z j_+,\label{fluxp}\\
\partial_z T_{zz}&=0, \, \label{eq:Tzz}
\end{align}
From~(\ref{eq:condvandfluxes})-(\ref{eq:condfluxes}), we obtain $\partial_z (v_n + j_s+j_++j_-)=0$ and $\partial_z (j_+-j_-)=0$ which we can integrate imposing the no-flux condition at $z=0$ to get:
\begin{align}
    j_+&=j_-,\label{eq:jconst}\\
    v_n &= - j_s-2j_+.\label{eq:vn2}
\end{align}
We can use~(\ref{eq:jconst}) to eliminate the electric potential $\Phi$ from the model, which can be obtained by solving
\begin{equation}
    \partial_z \Phi = \partial_z \left(\G p +\ln \phi_--\chi\phi_s\phi_n+\phi_n\right) +\frac{j_+}{\phi_-\D}-\frac{j_s}{\phi_s\D}, \label{ODEPhi}
\end{equation}
coupled with the Dirichlet boundary condition~(\ref{bc:Phi}).
\end{subequations}
Similarly we eliminate the pressure from the model by integrating $\partial_z (T_{zz})=0$ and applying the boundary condition~(\ref{cond:nostress}), which leads to
\begin{align}
 p =  \frac{\omega^2}{\G}\left[\phi_s\partial_{zz}\phi_s-\frac{(\partial_z
\phi_s)^2}{2}\right]+
\frac{\left(1-\phi_n^2\right)}{\phi_n}\, , \label{gov_b_pressure}  
\end{align}
The evolution of the gel composition is therefore dictated by the following two governing equations:
\begin{subequations}
\label{gov_a}
\begin{align}
    \partial_t \phi_s+\partial_z(\phi_s v_n) = -\partial_z j_s,\\
    \partial_t \phi_+ +\partial_z(\phi_+ v_n)= -\partial_z j_+.
\end{align}
\end{subequations}
which is coupled to the constitutive laws~(\ref{constlaws}) 
\begin{subequations}
\label{gov_b}
\begin{align}
j_s&=  - \frac{\phi_s^2}{1-\phi_n}\partial_z \mu_s +\frac{2\phi_s}{(1-\phi_n)\D}\,j_+\label{js},\\
j_+ &=- \frac{\D\phi_+\phi_-}{\phi_++\phi_-} \partial_z \bar{\mu}+ \frac{2\phi_+\phi_-}{\phi_s(\phi_++\phi_-)}\,j_s,\label{jpm}\\
v_n &= - j_s-2j_+,\label{vn}\\
\bar{\mu} &=A(\phi_s,\phi_+)+ 2\omega^2\phi_s \partial_{zz} \phi_s
-\omega^2\left(\partial_{z} \phi_s\right)^2,\label{mupmdef}
\\ 
\mu_s &= B(\phi_s,\phi_+) 
 -\left(1-\phi_s\right) \omega^2\partial_{zz} \phi_s
-\frac{\omega^2}{2}\left(\partial_{z} \phi_s\right)^2,
\label{musdef}\\
\phi_n &= \frac{1-\phi_s-2\phi_+}{1+\alpha_f},\label{phin}\\
\phi_- &=  \frac{\phi_++(1-\phi_s-\phi_+)\alpha_f}{1+\alpha_f}\label{eq:phim}.
\end{align}
Here, the functions $A$, $B$ are defined as follows:
\begin{gather}
\begin{aligned}
A(\phi_s,\phi_+)=\ln \left(\phi_+\frac{\phi_++(1-\phi_s-\phi_+)\alpha_f}{1+\alpha_f}\right) +2\left[1-\chi\phi_s\right]\frac{1-\phi_s-2\phi_+}{1+\alpha_f}\\
+\frac{2\G}{1+\alpha_f}\frac{(1+\alpha_f)^2-(1-\phi_s-2\phi_+)^2}{1-\phi_s-2\phi_+},
\label{eq:A}
\end{aligned}\\[2pt]
\begin{aligned}
B(\phi_s,\phi_+)= \ln \phi_s +\left[\chi(1-\phi_s)+1\right] &\frac{1-\phi_s-2\phi_+}{1+\alpha_f}\\
+&\frac{\G}{1+\alpha_f}\frac{(1+\alpha_f)^2-(1-\phi_s-2\phi_+)^2}{1-\phi_s-2\phi_+}.\label{eq:B}
\end{aligned}
\end{gather}
\end{subequations}
In some of the numerical simulation in the following sections, we also consider the scenario in which the gel is isolated from the bath. When this is the case, the boundary conditions~(\ref{muscont})-(\ref{mupmcont}) are replaced by no-flux conditions at the free interface (since mobile species are trapped in the gel). Moreover, (\ref{bc:Phi}) is replaced with a grounding condition at $z = h(t)$. Hence, for an isolated bath we impose
\begin{subequations}
\begin{align}
    j_s(h(t),t)=j_+(h(t),t)=0,\\
     \Phi(h(t),t)=0.\label{bc:nofluxb}
\end{align}\label{bc:noflux}%
\end{subequations}
Unless otherwise stated, the parameters used in the simulations are 
\begin{subequations}\label{par_sim2}
\begin{align}
\alpha_f&=0.04,\,\, \D=5,\,\, \chi=0.78,\,\,\G=2\times 10^{-4},\,\, \omega=0.025. 
\end{align}
We further consider the fraction $\phi_0$ to change at time $t=0$ from $\phi_{0^{-}}=10^{-6}$ to a new value $\ionbp$. For the latter we consider two scenarios:
\begin{align}
\mbox{set 1:}\quad	\ionbp &= 10^{-4}\quad\mbox{and}\quad\mbox{set 2:}\quad\ionbp = 10^{-2}.
\end{align}
\end{subequations}
The details on the numerical methods used to solve the system~(\ref{gov_a})-(\ref{gov_b}) in the different scenarios can be found in the supplemental material Section~\ref{numerics}. We note that the rescaled spatial variable which maps the gel to a fixed domain is  denoted by $Z=z/h(t)\,\in[0,1]$.

\begin{figure}[tb]
\centering
\begin{subfigure}{0.45\textwidth}
	\centering
	\includegraphics[width=0.9\textwidth]{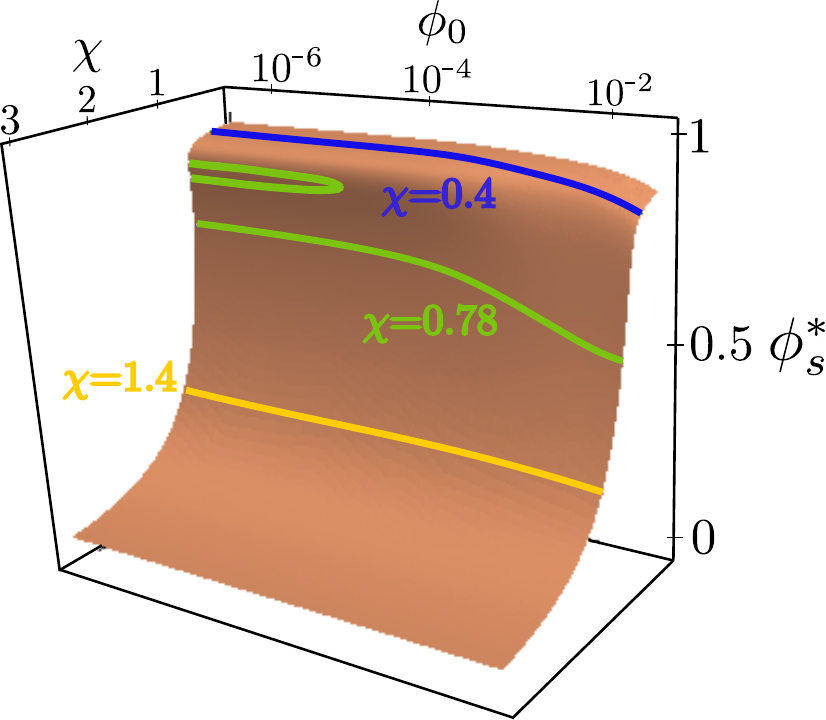}
	\caption{$\G=2\times 10^{-4}$}
	\label{3dmanifold}
\end{subfigure}
\begin{subfigure}{0.45\textwidth}
	\centering
	\includegraphics[width=0.9\textwidth]{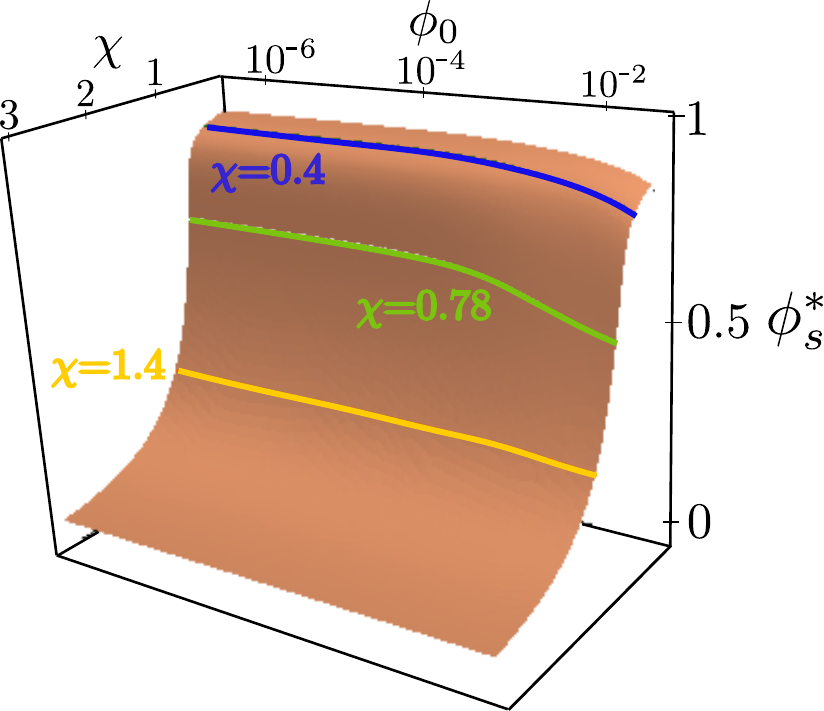}
	\caption{$\G=10^{-3}$}
	\label{3dmanifold2}
\end{subfigure}

\vspace{-5mm}
\caption{
Equilibrium manifold in the phase diagram $(\phi_0,\chi,\phi^*_s)$
for parameters as given in equation (\ref{par_sim2}). 
The curves show sections of the manifold at specific values of $\chi$.
}
\end{figure}
\section{Homogeneous equilibrium states and routes to collapse}
\label{sec_eq}

\begin{figure}[p]
\centering
	\includegraphics[width=0.95\textwidth]{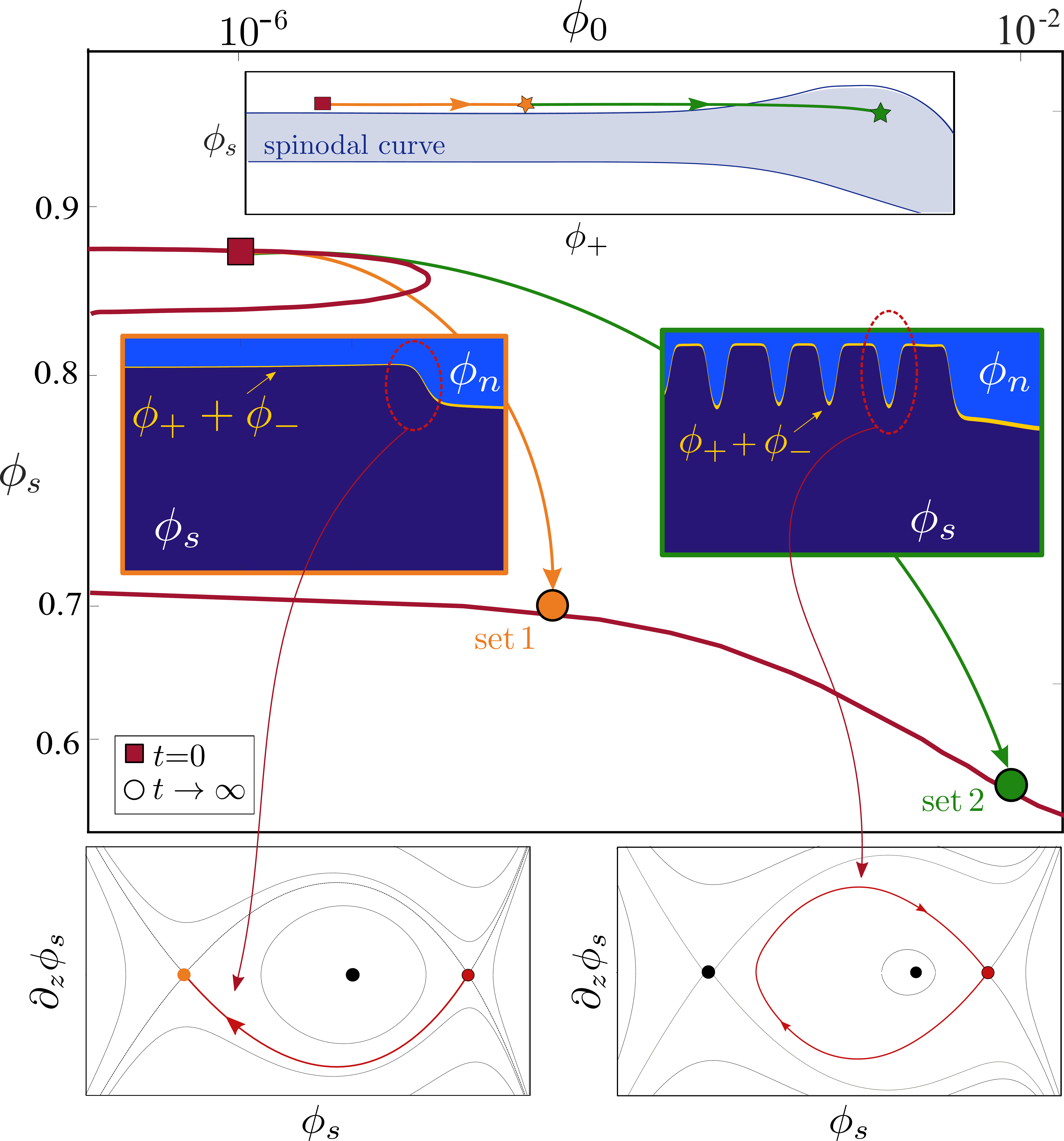}
	\caption{
{\bf Routes to collapse}: Equilibrium curves $\phi_s(\phi_0)$ (same as the green curve in Fig.~\ref{3dmanifold}). The collapse is initiated by increasing $\phi_+$ at $t=0$ (red square) towards a value (orange star) outside the spinodally unstable domain (blue region in top panel) and another value inside the spinodally unstable domain (green star). The two scenarios are shown by the corresponding trajectories (orange and green), where in the first case,  collapse occurs via a depletion front (set 1) and in the second case via a depletion front that is accompanied by phase separation via spinodal decomposition (set 2). Phase-plane analysis (bottom) predict heteroclinic orbits connecting the two equilibria (left) and  homoclinic orbits in the latter case (right).
}
\label{equi1_stab}
\end{figure}
We now consider the system (\ref{gov_a})--(\ref{gov_b}) for $\phi_s$ and $\phi_+$ in equilibrium with an ionic bath. More specifically, we consider homogeneous, flux-free steady states that satisfy the boundary conditions  \eqref{bottombc}--\eqref{bbc}. 
The equilibrium states are denoted by 
$(\phi^*_{s},\phi^*_{+})$ and 
satisfy
\begin{subequations}\label{homb}
\begin{align}
A(\phi^*_s,\phi^*_+)-2\ln(\phi_0)&=0,\label{Feqpm}\\
B(\phi^*_s,\phi^*_+)-\ln(1-2\phi_0)&=0. \label{Feqs}
\end{align}
\end{subequations}
Recall that $\phi_0$ denotes the volume fraction of ions in the surrounding bath.
The conditions \eqref{homb} define the generalisation of the Donnan equilibrium \cite{donnan_theory_1924}
as it applies to our situation, i.e. a polyelectrolyte gel with its fixed charges and the mobile species represented by the salt ions~\cite{huyghe_quadriphasic_1997}.

Considering the shear modulus of the gel $\G$ to be fixed, we investigate how the equilibria depend on the concentration of ions in the bath $\phi_0$ and the Flory interaction parameter $\chi$. For large shear moduli $\G$, as in Fig.~\ref{3dmanifold2}, the system presents a unique stable steady state. While  variations of $\phi_s^*$ along $\phi_0$ are small, there is a sensitive dependence on the Flory interaction  parameter $\chi$.
For small $\G$, as in Fig.~\ref{3dmanifold}, the system undergoes a bifurcation, where a regime with three equilibrium states moves to one with a single equilibrium state as  $\chi$ or $\phi_0$ are varied. The stability properties of these equilibria will be characterized in Section \ref{sec_stability}. At this point we note that, 
as shown in Fig.~\ref{equi1_stab}, for high $\phi_0$ the only steady state is characterised by a low concentration of solvent, i.e. a collapsed state; as we decrease the concentration $\phi_0$, an additional disconnected branch appears. The latter is characterised by two highly swollen steady states, a stable (upper branch) and an unstable one (middle branch). In other words the system undergoes a saddle-node bifurcation.
\paragraph{Routes to collapse}
 As shown in Fig. \ref{equi1_stab} (set 1), the volume transition is accompanied by the development of a depletion front which separates two homogeneous states, the highly swollen region away from the free interface (gel bulk) and the poorly swollen region near the moving boundary $z=h(t)$. We will predict this using a phase-plane analysis in the next section \ref{sec_phaseplane}, where we show how the front can be approximated by quasi-steady, non-homogeneous solutions of the full model. Another scenario arises if the ion concentration is increased further. It results in phase separation in the bulk of the gel and the formation of regions of high and low solvent concentration, the latter showing also higher concentration of ions. A linear stability analysis to discuss spinodal decomposition, given in Section \ref{sec_stability}, investigates the second scenario, where for sufficiently large  $\ionbp$ (such as for set 2), the propagation of the depletion front occurs after the spinodal decomposition of the gel. As shown in top panel of Fig.~\ref{equi1_stab}, the bulk decomposition is initiated by the rapid increase in $\phi_+$ in the gel (see green arrow in top panel) which drives the system into the region of instability (blue region enclosed by the spinodal curve).

\section{Front propagation in a collapsing gel}\label{sec_collapse}
\begin{figure}[t]
	\begin{subfigure}{0.32\textwidth}
		\hspace{-5mm}		
		\includegraphics[width=\textwidth]{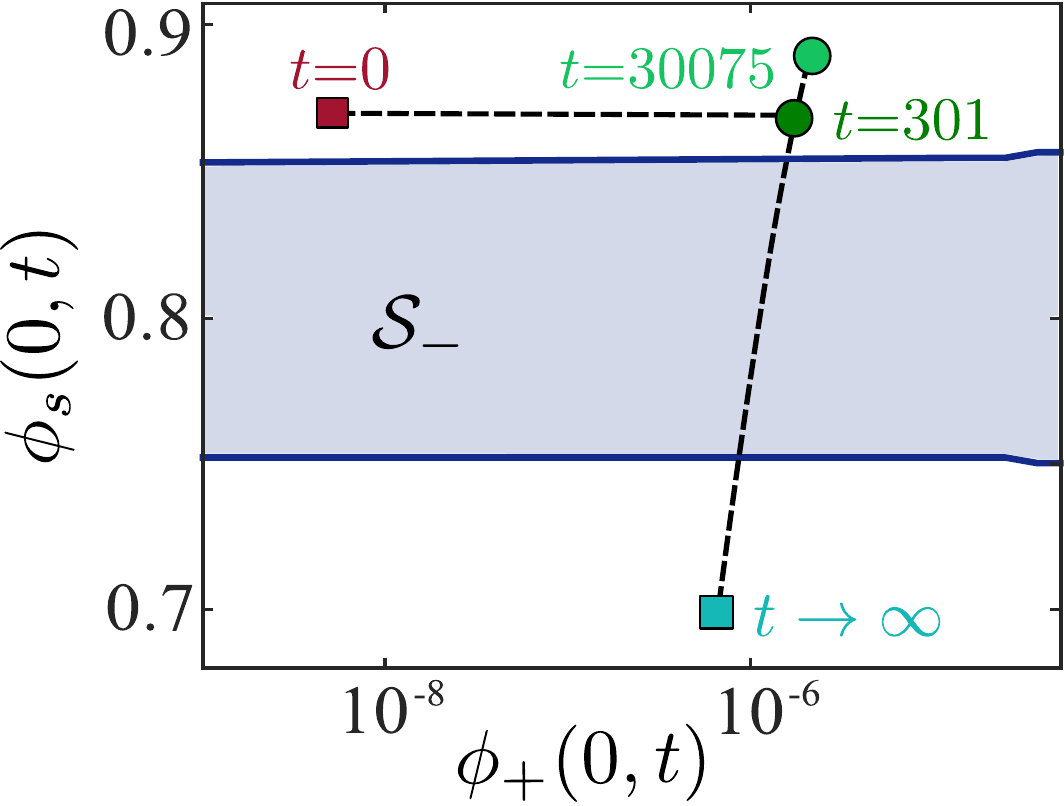}
		\caption{}
		\label{figset1:phaseplane}
	\end{subfigure}
\hspace{1mm}
	\begin{subfigure}{0.32\textwidth}	
		\hspace{-5mm}
		\includegraphics[width=\textwidth]{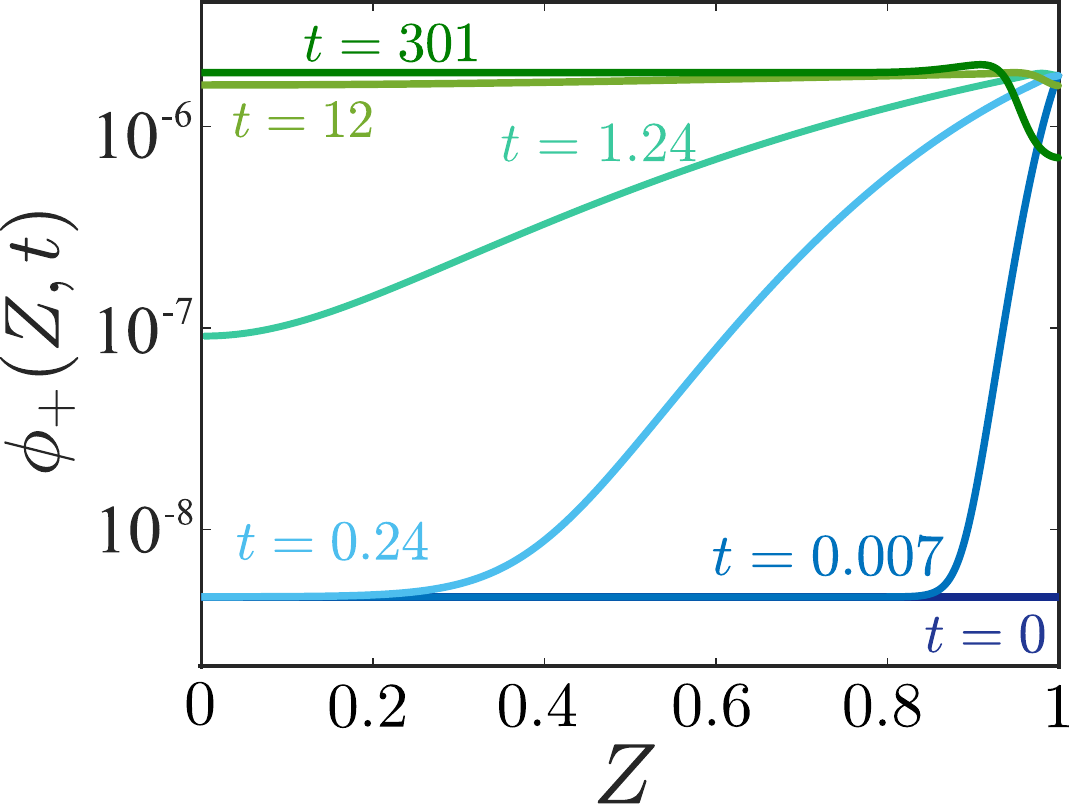}
		\caption{}
		\label{figset1:earlyphip}
	\end{subfigure}
\hspace{1mm}
	\begin{subfigure}{0.32\textwidth}	
		\hspace{-6mm}
	\includegraphics[width=\textwidth]{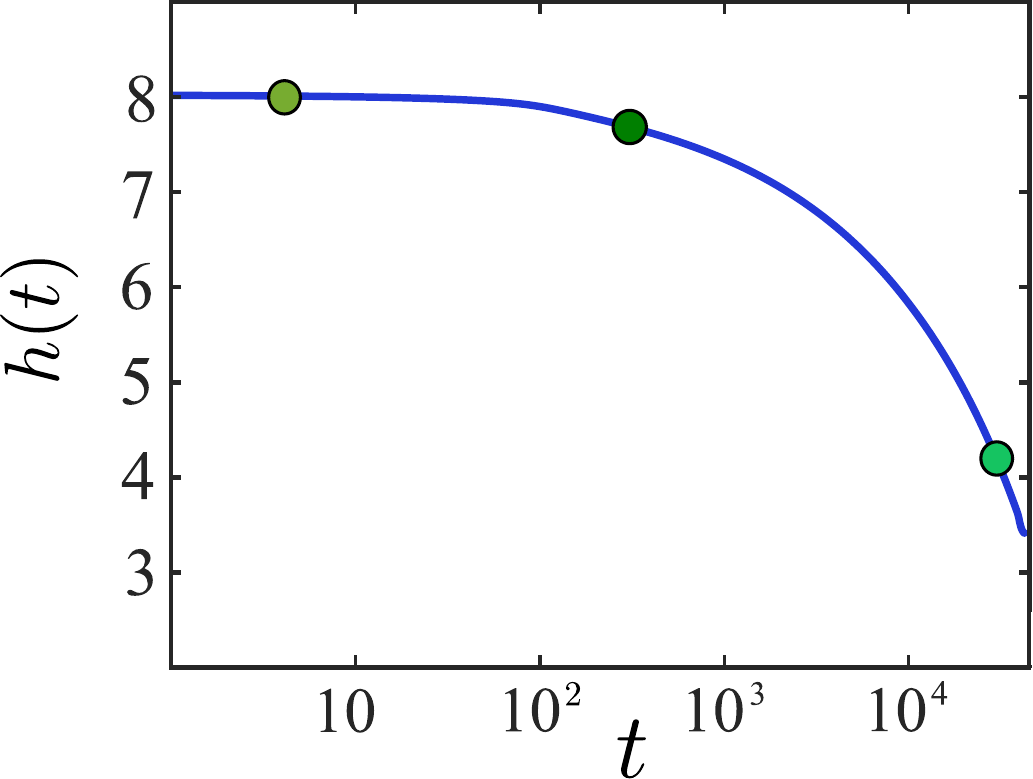}
	\caption{}
	\label{figset1:h}
	\end{subfigure}

	\vspace{-5mm}
	\caption{Numerical solutions for the case of a dilute bath for the parameters set 1 in (\ref{par_sim2}).
          (a) Evolution of the solution at the substrate (i.e. $Z=0$) in the $(\phi_s,\phi_+)$ plane. The instability region $\mathcal{S}_-$
          as predicted by
		the stability analysis is highlighted in blue. (b) Evolution of the ion
		fraction $\phi_+(Z,t)$. (c) Time evolution of the size of the gel (colour of the dots corresponds to those used in (a) and (b)).
	}\label{figsimset1}
\end{figure}

\begin{figure}[tb]
	\hspace{-2mm}
	\includegraphics[width=\textwidth]{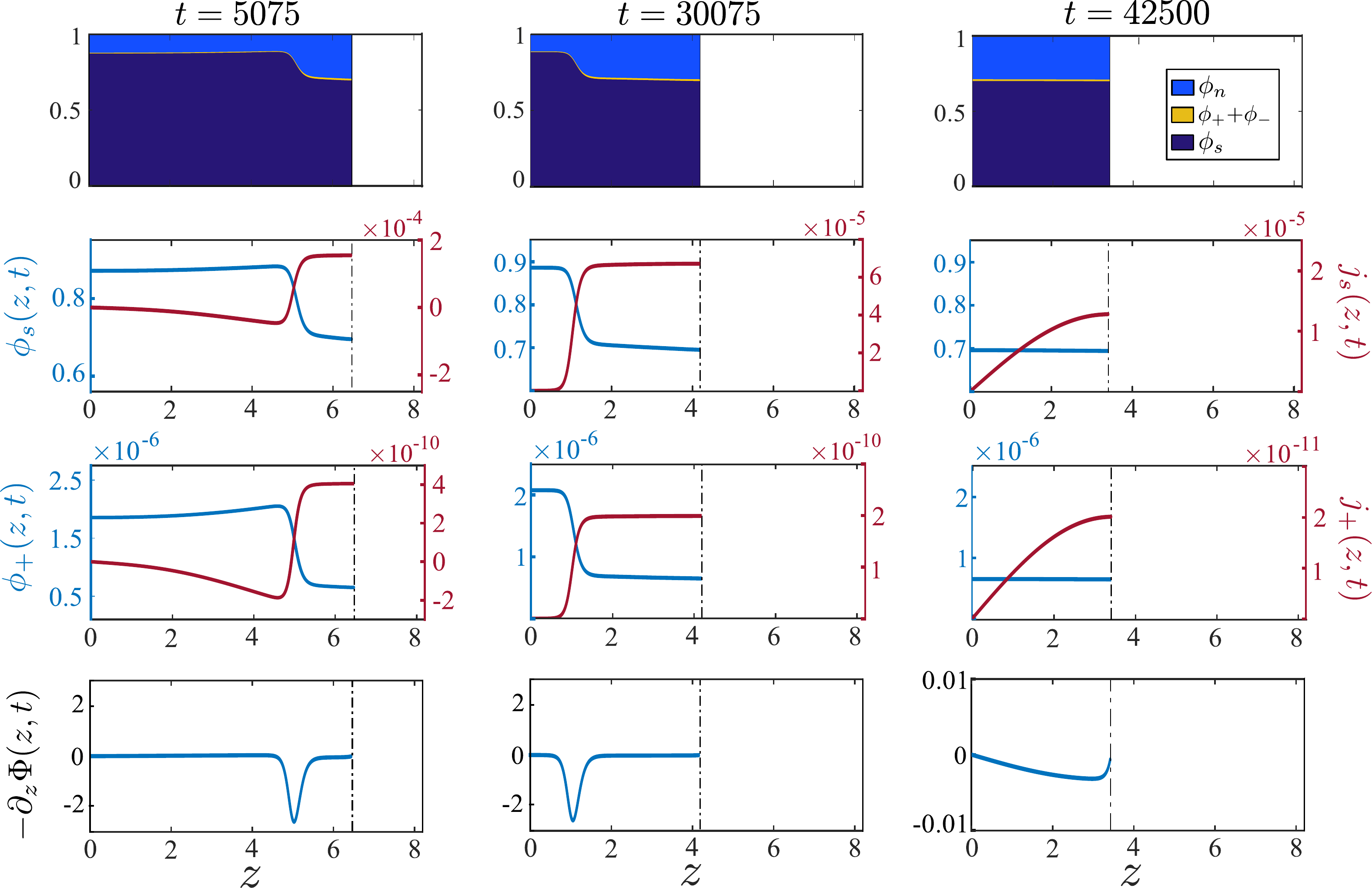}	
	\caption{Numerical solution of the model~(\ref{gov_a})-(\ref{gov_b}) for parameter set 1 in (\ref{par_sim2}) and boundary conditions~(\ref{bottombc})-(\ref{bbc}) (same scenario as Fig.~\ref{figsimset1}): The first row illustrates the gel composition at different point in time. For the same time points we also plot the volume fractions, the corresponding fluxes 
	and the electric field $-\partial_z \Phi$ (defined by using~(\ref{ODEPhi})).}
	\label{set1A}
\end{figure}

For the scenarios described above, the dynamics of the ions occur on a
much faster time scale than those of the solvent. 
As shown in Fig.~\ref{figset1:phaseplane}, 
for $t\sim O(1)$, the ion concentration quickly increases while $\phi_s$ remains approximately constant. As the gel is initially highly swollen, i.e. $\phi_+ \ll \phi_s$, the gel size is determined by the concentration of solvent, and so remains approximately constant during this first transient (see Fig. \ref{figset1:h}), i.e. $h(t)\approx h(0)$ for $t\sim O(1)$ with $h(0)\approx (1-\phi_s(0))^{-1}\approx 8$.
Focusing on the evolution of $\phi_+$ (see Fig.~\ref{figset1:earlyphip}), we see that within less than one time unit, the
concentration builds up at the boundary and subsequently penetrates into the gel.
At $t=12$,
the process has almost concluded and in fact, early signs of a new front 
manifest themselves at the free interface, which becomes more pronounced at
$t=301$. Together with the ion concentration, the generalised chemical potential
$\bar{\mu}$ move from its initial value $2\ln(\phi_{0^{-}}))$ to approximately
$2\ln(\ionbp)$.  
This difference in chemical potential drives the process of ion
diffusion. Since the flux of ions is $\D$ times the gradient of the chemical
potential (and the domain size $h(0) \approx 8$), we can estimate the time scale
  of ionic diffusion as $t \sim [h(0)]^2/ [ 2 \D \ln(\ionbp/\phi_{0^-})]
  \approx 2.15$, which is consistent
with the observation in the numerical simulations. Using a scaling argument we can also estimate the concentration of ions in the gel prior and after the transient.
Given that the concentration of ions in the bath is small, i.e. $\phi_0\ll 1$, the logarithmic term $\ln \phi_0$ in~(\ref{Feqpm}) becomes large and needs to be balanced. Since the gel is not too swollen, i.e. $1-\phi_s-2\phi_+\gg\phi_0$, or dry, i.e. $\phi_n^{-1}\G\ll 1$
the only term in \eqref{eq:A} that can balance $\ln \phi_0$ is the logarithmic term so that 
\begin{equation}
\phi_+\frac{\phi_++\alpha_f(1-\phi_s-\phi_+)}{1+\alpha_f}\sim \phi^2_0.\label{eq:balance1}
\end{equation}
and given the chosen value of $\alpha_f$ (see~\ref{par_sim2})), the balance~(\ref{eq:balance1}) gives 

\begin{equation}\label{estphipm}
\phi_+\sim \phi_0^2(1-\phi_s)^{-1}/\alpha_f,
\end{equation}
so that we predict the concentration of ions to be $\phi_+\sim 1.25 \times 10^{-10}$ prior to $t=0$ and $\phi_+\sim 1.25 \times 10^{-6}$ after the fast ion diffusion (given $(1-\phi_s)^{-1}\approx 8$). These are in good agreement with numerical result in Fig.~\ref{figset1:earlyphip}.
After the ion concentration has equilibrated, a slower process takes place,
whereby solvent is removed from the gel through the aforementioned 
depletion front. This is illustrated in Fig.~\ref{set1A}, where we present snapshots in time of the gel composition (see first row). For the same time points of the snapshots we also illustrate the values of the volume fractions, chemical potentials and electric field in the gel. The front is clearly seen in the concentration profiles for
the solvent $\phi_s$ as well as for the mobile ion species $\phi_+$ and the gradient of the electric potential $-\partial_z \Phi$, i.e. the electric field in the gel. In the poorly swollen region of the gel (near the free interface), $\phi_s\approx 0.7$; using the scaling~(\ref{estphipm}) we therefore expect reduction in the concentration of ions, more precisely we obtain $\phi_+\sim 8\times 10^{-7}$ which is again in line with the numerical results.
In  the last column ($t=42500$), the front has reached the
substrate and the gel has collapsed with its composition having reached a new
state (see `powder' blue square in Fig.~\ref{figset1:phaseplane}). The solvent flux $j_s$ is only on the order of $10^{-4}$, hence the appropriate time scale for
the depletion front movement is $O(10^4)$. The main contribution is from the gradient of the chemical potential $\mu_s$ which is set by the function $B$ (see Eq.~(\ref{eq:B})). The latter is connected with $\cal G$, which is small and it is this small value that determines the slow collapse. 

To summarize (as illustrated in Fig.~\ref{figset1:phaseplane}), the gel starts from a state of high concentration of solvent and a small concentration of co-ions. From there, it quickly evolves to a new state with higher salt
content. Both of these states are stable and show no sign of spinodal decomposition. Instead, a depletion front moves through the gel. At later time (i.e. $t\sim 30000$), as the front approaches the substrate, the solvent concentration $\phi_s(0,t)$ increases slightly before decreasing to its new linearly stable steady state value, with a slightly higher salt but a much lower solvent concentration. 

\subsection{Phase-plane analysis}\label{sec_phaseplane}
\begin{figure}[t]
	\includegraphics[width=0.9\textwidth]{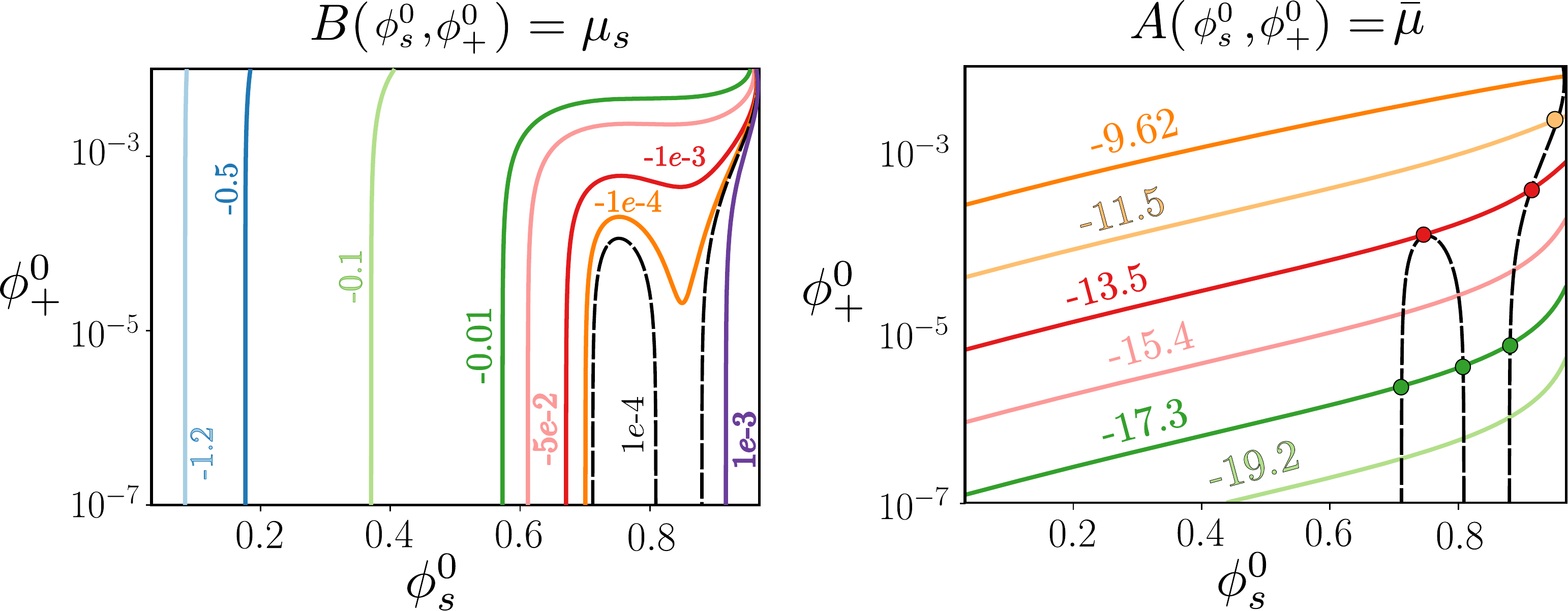}
	\caption{Level sets for the functions $A$ and $B$ that define the fixed points of the system~(\ref{qncode}), note that the functions are expressed in terms of $\phi_s^0$ and $\phi_+^{0}$. The dashed line in right panel is taken from the left one (i.e. $\mu_s\equiv 10^{-4}$).
	}
	\label{fig:levelset}
\end{figure}

\begin{figure}[t]
\includegraphics[width=\textwidth]{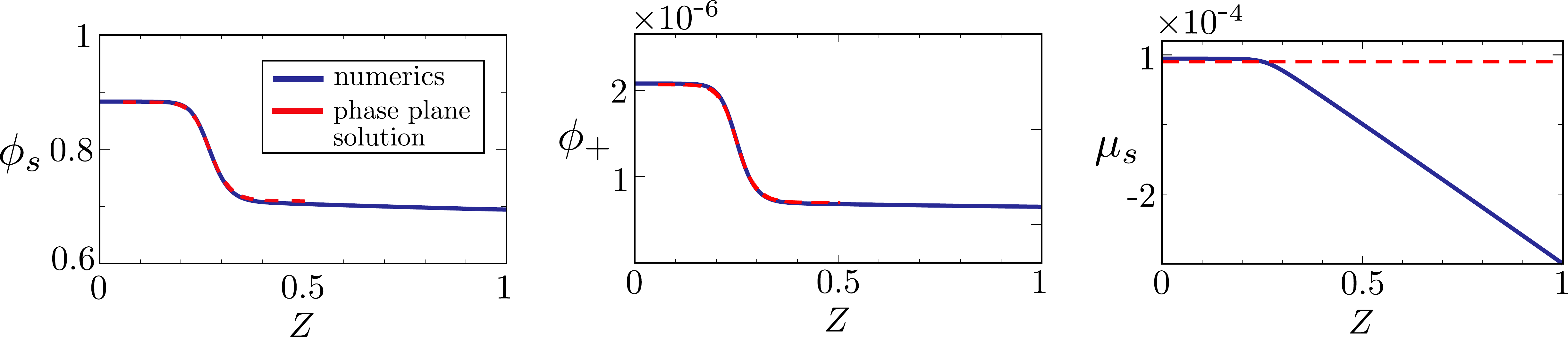}
\caption{Comparison of the asymptotic approximation for the phase-plane
solution \eqref{phiqphip-ode}--\eqref{maxwell} (shown with dashed red
lines) for $\phi_s$, $\phi_+$ and $\mu_s$ with the numerical results from Fig.~\ref{set1A} at time $t=30075$ (solid black lines), using the parameter set 1 in Table~\ref{par_sim2}.  A single shift along the $Z$-axis was applied to all phase plane profiles so that
$\phi_s$ matches the numerical solution at $\phi_s=0.8$.
}\label{odepdefig}
\end{figure}

\begin{figure}[ht]
	\includegraphics[width=0.95\textwidth]{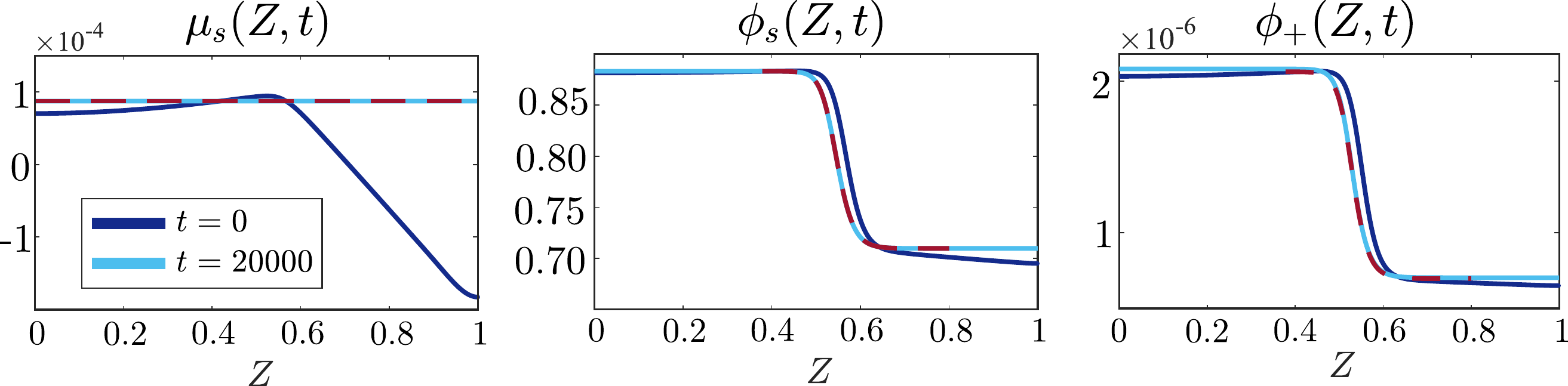}
	\caption{Spatial distribution of the variables at different times.  We
		start with the results from the simulations in
		Fig.~\ref{set1A} at time $t=15075$.  This provides the initial conditions for
		the results shown here. The gel is taken out of the bath allowing the system to relax towards a non-homogeneous
		steady state 
		which is compared to the
		prediction from the phase plane analysis (dashed line).} \label{nofluxset1A}
\end{figure}

The slow movement of the depletion fronts on the diffusive time scale 
suggests that these structures are in a quasi-stationary state, which simplifies  their analysis by enabling the time derivatives and 
fluxes from the system~(\ref{gov_a})-(\ref{gov_b}) to be neglected, so that the solution will only depend on the spatial variable $z$. Since fluxes are negligible, $\mu_s$, $\bar{\mu}$ are almost constant (independent of $z$) in an
$O(1)$ vicinity of the front.
With this assumption and after rescaling $z$ with $\omega$, we have,
\begin{subequations}\label{qncode}
\begin{align}
\partial_z \phi_s&=q\\
\partial_z q &= 
\frac1{1-\phi_s}\left[B(\phi_s,\phi_+)-\mu_s\right]-\frac1{2(1-\phi_s)}q^2
\label{qncode-q}\\
\mu_s&=\frac{\phi_s-1}{2\phi_s} \left[\bar{\mu}-A(\phi_s,\phi_+)\right]
+B(\phi_s,\phi_+)-\frac1{2\phi_s}q^2,
\label{consqnc}
\end{align}
\end{subequations}
where $A$ and $B$ are as defined by~(\ref{eq:A})-(\ref{eq:B}), and we introduce the auxiliary variable $q$. 
The fixed points $(q,\phi_s,\phi_+)=(0,\phi_s^0,\phi_+^0)$ of the system (which correspond to homogeneous equilibrium states for the full model) are found from \eqref{qncode} via
\begin{equation}
\label{flat}
A(\phi_s^0,\phi_+^0)=\bar{\mu} ,\quad
B(\phi_s^0,\phi_+^0)=\mu_s.
\end{equation}
As shown in Fig.~\ref{fig:levelset} the number of fixed points varies depending on the values assigned to $\mu_s$ and $\bar{\mu}$. For sufficiently small values of $\mu_s$ (see $\mu_s=-0.1$ for example) the system has a unique fixed point. As we increase the value of $\mu_s$ there is an intermediate region near $\mu_s=0$ where multiple fixed points can exist depending on the values of $\bar{\mu}$. As shown on the right-hand side panel of Fig.~\ref{fig:levelset}, when considering $\mu_s=10^{-4}$, the system has three fixed points when $\bar{\mu}$ is below a critical negative value, i.e. in the dilute limit $\phi_+^0\ll 1$. As $\bar{\mu}$ increases, the system undergoes a saddle-node bifurcation resulting in a unique fixed point. Note that the equilibrium states computed in Section~\ref{sec_eq} are a subset of the fixed points of the system \eqref{qncode} as the value of $\mu_s$ and $\bar{\mu}$ are not independent as they are set by the ionic bath via the boundary conditions~(\ref{muscont})-(\ref{mupmcont}).
For our local analysis we note that the depletion front represents a narrow, slowly moving transition between two 
almost homogeneous states, so that we can approximate it 
by a quasi-stationary solution that tends to homogeneous states
as $z\to\pm \infty$. We then linearise around $(0,\phi_s^0,\phi_+^0)$
and determine the number of modes consistent with these
limits, which we then use to carry out a degree of freedom count.
We make the ansatz
\begin{equation}
\phi_s=\phi_s^0+\delta\phi_s^1 \mathrm{e}^{s z}, \quad
q=\delta q^1 \mathrm{e}^{s z}, \quad
\phi_+=\phi_+^0+\delta \phi_+^1  \mathrm{e}^{s z}, 
\end{equation}
with $\delta\ll 1$, and $s$ being a constant. Inserting this, we obtain at $O(\delta)$ the condition
\begin{equation}
\begin{bmatrix}
B_{\phi_s}-(1-\phi_s^0) s^2&
B_{\phi_+}\\
&\\
(1-\phi_s^0)A_{\phi_s}+{2\phi_s^0}\,B_{\phi_s}&
(1-\phi_s^0)A_{\phi_+}+{2\phi_s^0}\,B_{\phi_+}
\end{bmatrix}
\begin{bmatrix}
\phi_s^0\\ \\ \phi_+^0
\end{bmatrix}
=0,
\end{equation}
where the subscripts $\phi_s$ and $\phi_+$ denote partial derivatives of $A$ and $B$, which are evaluated at the equilibrium point $(\phi_s^0,\phi_+^0)$. In order for the system to have non-trivial solutions, we must set the determinant of the coefficient matrix to zero, which gives
\begin{equation}\label{ssquare}
s^2=
\myfrac[2pt]{A_{\phi_+}  B_{\phi_s} - A_{\phi_s}   B_{\phi_+}}{\left(1-\phi_s^0\right) A_{\phi_+}+2\phi_s^0 \,B_{\phi_+}}.
\end{equation}
We therefore conclude that the equilibrium is a  saddle point if $s^2>0$. 
We now seek a non-homogeneous quasi-stationary solution connecting two saddle points,  $(0,\phi_s^a,\phi_+^a)$ and  $(0,\phi_s^b,\phi_+^b)$. This
corresponds to imposing the far-field conditions $(\phi_s,\phi_+)\to(\phi_s^a,\phi_+^a)$ and $(\phi_s,\phi_+)\to(\phi_s^b,\phi_+^b)$
as $z\to-\infty$ and $z\to+\infty$, respectively. 
When considering the degrees of freedom in the solution, 
we have one mode for each saddle point which is consistent with the limit, 
 namely $s^a<0$ for $(0,\phi_s^a,\phi_+^a)$ and $s^b>0$ 
for $(0,\phi_s^b,\phi_+^b)$.
These account for two degrees of freedom, which, 
together with the two unknown constants $\mu_s$ and $\bar{\mu}$, add up to
four degrees of freedom. 
The system \eqref{qncode} is second order and hence removes two degrees of freedom, and
the invariance of any solution with respect to translations along the $z$-axis
subtracts
another one. Hence one degree of freedom remains, and this is associated with 
$\bar{\mu}$, which needs to be given. We note that a full expansion at higher orders about the quasi-stationary solution might lead to additional constraints that fixes the degree of freedom remaining in the system. However, this goes beyond the purpose of this analysis, which aims at understanding the structure of the depletion front. 
\paragraph{The limit of dilute salt concentrations} 
We here consider the limit of dilute salt concentration, i.e. $\phi_+\ll\phi_s<1$, which is representative of the numerical results from Fig.~\ref{set1A}, where $\phi_+$ is indeed small. Inspired by the scaling~(\ref{estphipm}), we 
rescale the model variables, via
\begin{equation}
\phi_+=\alpha_f \epsilon^2\tilde\phi_+,
\quad
\bar{\mu}=2\ln(\alpha_f \epsilon)+\tilde\mu.\label{eq:dil_limit}
\end{equation}
where the dilute limit is therefore taken by considering $\epsilon \rightarrow 0$. Note that if we define $\epsilon=\phi_0/\alpha_f$, we recover~(\ref{estphipm}) up to a factor $1-\phi_s$ allowing us to relate the phase plane analysis to the numerical simulations in the presence of a bath (which is instead here neglected).
Substituting~(\ref{eq:dil_limit}) into~(\ref{qncode}) and considering only the leading order problem in this limit, we obtain
\begin{subequations}\label{phiqphip-ode}
\begin{align}
\partial_z \phi_s&=q \label{phiqphip-odea}\\
\partial_z q &= 
\frac1{1-\phi_s}\left[B_0(\phi_s)- \mu_s\right]-\frac1{2(1-\phi_s)}q^2\label{phiqphip-odeb}\\
\mu_s&=\frac{1-\phi_s}{2\phi_s} \left[A_0(\phi_s,\tilde\phi_+)-\tilde\mu\right]
+B_0(\phi_s)-\frac1{2\phi_s}q^2,\label{phiqphip-odec}
\intertext{with}
A_0(\phi_s,\tilde\phi_+)&=
\ln\left[\left(\frac{1-\phi_s}{1+\alpha_f}\right)\frac{\tilde\phi_+}{\phi_s^2}\right]
-2\chi\left(\frac{1-\phi_s}{1+\alpha_f}\right)
+2B_0(\phi_s),\\
B_0(\phi_s)&=\ln(\phi_s)+(\chi(1-\phi_s)+1)\left(\frac{1-\phi_s}{1+\alpha_f}\right)
+{\cal G}\frac{(1+\alpha_f)^2-(1-\phi_s)^2}{(1+\alpha_f)(1-\phi_s)}.
\end{align}
\end{subequations}
\begin{figure}[t]
	\hspace{2mm}
	\begin{subfigure}{0.3\textwidth}
		\hspace{-4mm}
		\includegraphics[width=0.95\textwidth]{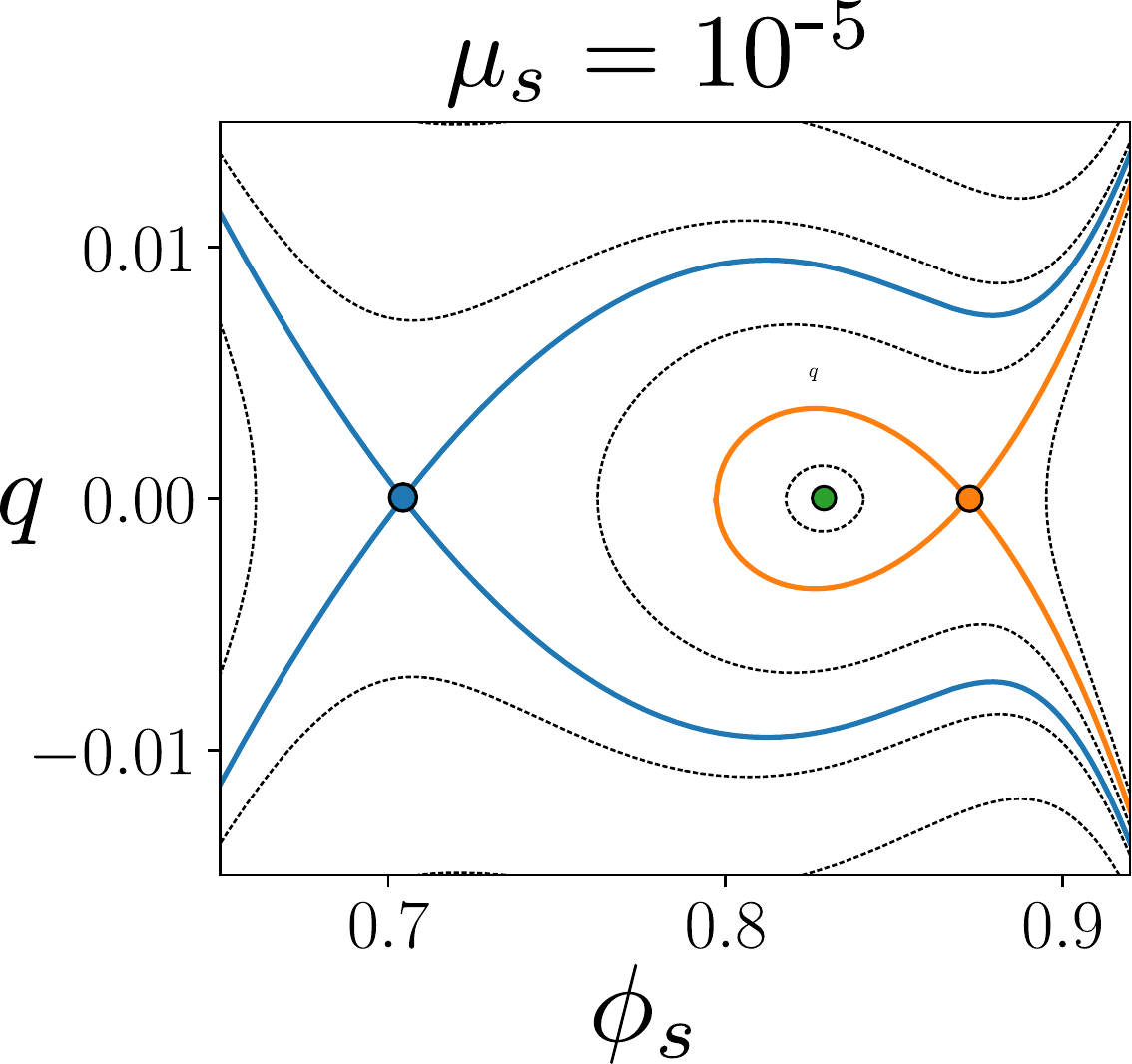}
		\caption{}
		\label{fig:dilute_PP1}
	\end{subfigure}	
	\begin{subfigure}{0.3\textwidth}
		\hspace{-4mm}
		\includegraphics[width=0.95\textwidth]{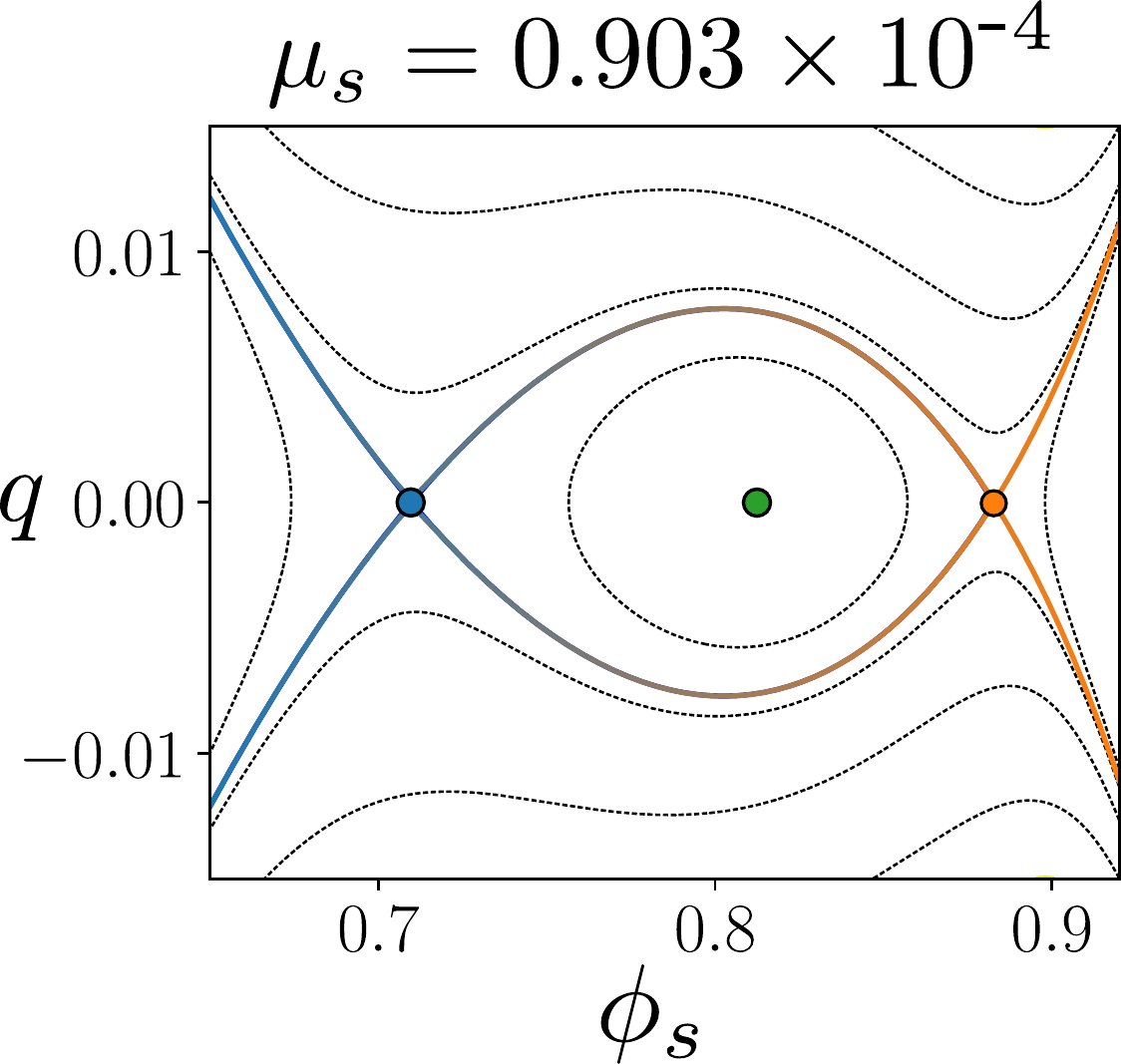}
		\caption{}
		\label{fig:dilute_PP2}
	\end{subfigure}	
	\begin{subfigure}{0.3\textwidth}
		\hspace{-4mm}
		\includegraphics[width=0.95\textwidth]{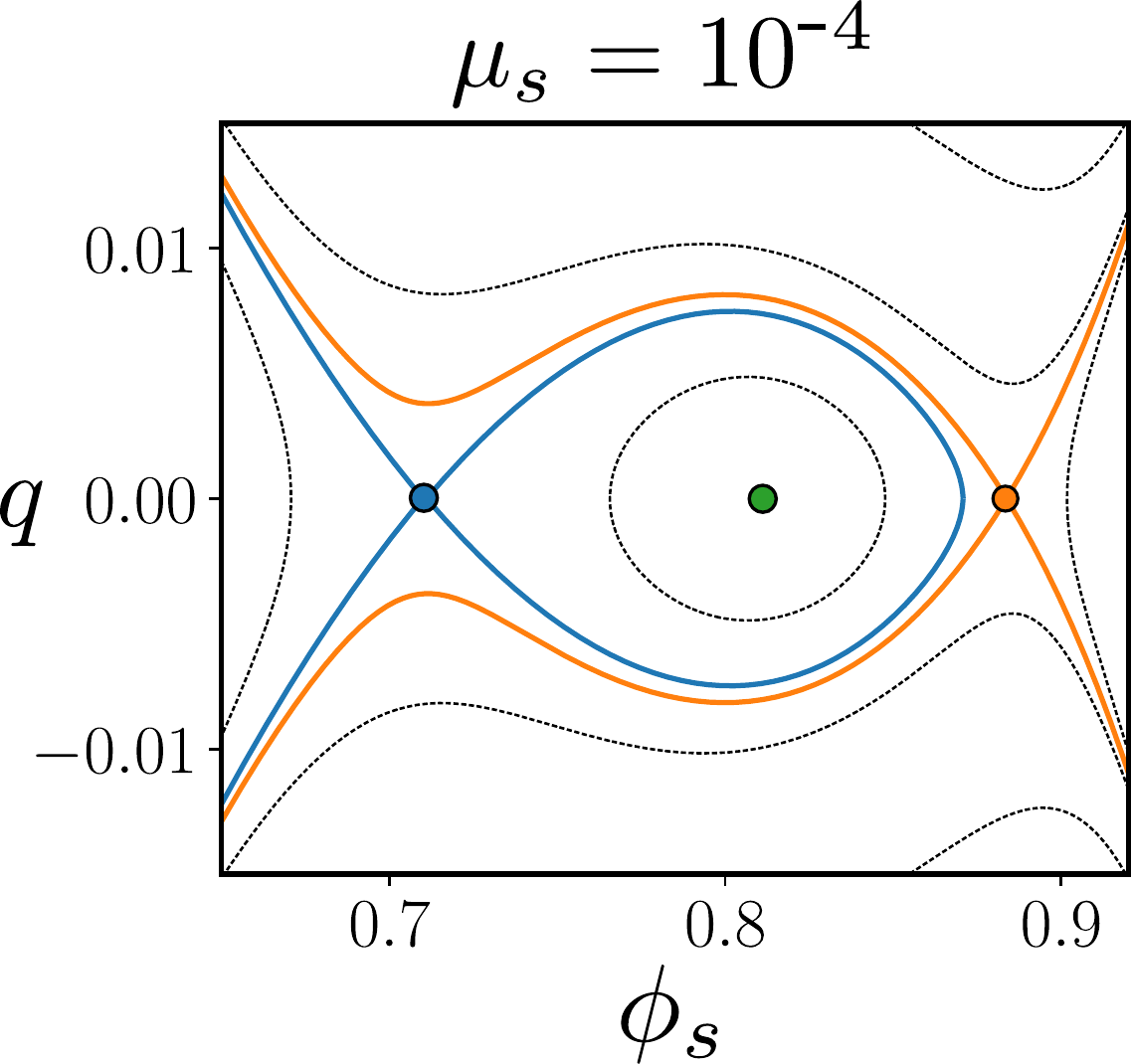}
		\caption{}
		\label{fig:dilute_PP3}
	\end{subfigure}	

	\vspace{-4mm}
	\caption{Phase plane analysis for the dilute limit for three different value of $\mu_s$: (a) $\mu_s<\mu_s^{(c)}$; (b) $\mu_s=\mu_s^{(c)}$ for which there is an heteroclinic orbit connecting the two equilibrium state; (c) $\mu_s>\mu_s^{c}$. Fixed point are highlighted by small circles, trajectories leaving from or asymptoting into fixed points are indicated by solid coloured line, while all other trajectories are denoted by dashed black line.}
	\label{fig:dilute_PP}
\end{figure}
Equations \eqref{phiqphip-odea} and \eqref{phiqphip-odeb} decouple from the 
algebraic constraint \eqref{phiqphip-odec}.  The first integral of \eqref{phiqphip-odea} and \eqref{phiqphip-odeb} is then used to investigate the phase plane as a function of the chemical potential $\mu_s$ as illustrated in Fig.~\ref{fig:dilute_PP}. In contrast to a non-ionic hydrogel \cite{Hennessy2020}, the contribution of the fixed charges allows multiple fixed points to exist when $\mu_s=0$, i.e. for a gel in contact with pure water. There is a unique critical value of $\mu_s$, denoted by $\mu_s^c$, for which the two saddle points are connected by an heteroclinic orbit, corresponding to a front-type non-homogeneous quasi-stationary solution (see Fig. \ref{fig:dilute_PP2}). If $\mu_s<\mu_s^c$ as shown in Fig.~\ref{fig:dilute_PP1}, there is a homoclinic orbit attached to the right fixed point (i.e.\ the orange point), corresponding to a quasi-stationary solution with
a localised solvent-depleted zone. 
On the contrary, for $\mu_s>\mu_s^c$, the homoclinic orbit is attached to the left fixed point (i.e.\ the blue point; see Fig.~\ref{fig:dilute_PP3}), which represents a localised solvent-rich zone. 
We can find the critical value of $\mu_s$ by combining \eqref{phiqphip-odea} and \eqref{phiqphip-odeb} into a second-order equation and using $f=1/\sqrt{1-\phi_s}$ as an integrating factor to obtain
\begin{subequations}\label{maxwell}
\begin{align}
\int_{\phi_s^a}^{\phi_s^b} \frac{B_0(\phi_s)-\mu_s}{(1-\phi_s)^2}
\d \phi =0,\label{intc}\\
B_0(\phi_s^a)-\mu_s=0, \quad B_0(\phi_s^b)-\mu_s=0,\label{fp}
\end{align}
\end{subequations}
where $\phi_s^a$ and $\phi_s^b$ are the values of the solvent fraction at the two saddle points. Note that this is a Maxwell condition
for the co-existing states $\phi_s^a$ and $\phi_s^b$, which are independent of the ion fraction $\tilde\phi_+$ in the dilute limit. 
Solving numerically the system with the parameters for $\alpha$ and $\chi$ and ${\cal G}$
corresponding 
parameter set 1 in Table~\ref{par_sim2} gives
\begin{equation}\label{mwvals}
\mu_{s} = 0.903\times 10^{-4},\quad 
\phi_s^a= 0.883, \quad
\phi_s^b= 0.710. 
\end{equation}
Note that provided $\tilde{\mu}$ is known from the numerical simulations, we can use the constraint~(\ref{phiqphip-odec}) to compute $\tilde\phi_+$.
Here
$\tilde\mu$ is small, on the order of $10^{-4}\ln(\epsilon\alpha_f)$, and we set them to be zero.  
As shown in Fig.~\ref{odepdefig}, the phase plane analysis is overall in good agreement with the dynamical simulation from Fig.~\ref{set1A}. While the value of $\mu_s$ at the free interface (i.e. $Z=1$) is set by the bath, in the bulk of the gel $\mu_s$ sets around the critical value obtained by our phase-plane analysis (see Eq.~(\ref{mwvals})). 
A setting that is more representative of the phase-plane analysis is a no-flux situation, where, on the long term, the fluxes vanish and the system can settle down to a non-homogeneous equilibrium.
(see light blue curve in Fig.~\ref{nofluxset1A} for long time behaviour). Unlike in Fig.~\ref{odepdefig} where there are visible gradients in the concentration
$\phi_s$ and $\phi_+$ even away from the front, in the no-flux case (see Fig.~\ref{nofluxset1A}) the solution profiles become exact stationary fronts that
converge to flat states away from the transition layer. The numerically estimated equilibrium values are $\mu_s\approx 0.8726\times 10^{-4}$, $\phi_s^a\approx0.8829$ and $\phi_s^b\approx0.7098$, which well agree with the values in~(\ref{mwvals}) if not for the deviation in the estimate of $\mu_s$. An even better approximation can be obtained by considering the general case \eqref{qncode}-\eqref{flat} (more details in the next section ) for which $\mu_s=0.873\times 10^{-5}$ being accurate
to within less than $0.1\%$ of this value in comparison to the dilute limit (see Eq. \eqref{mwvals}) which is accurate to within $3\%$.

\paragraph{The general case} 
We now explore the non-dilute limit. This requires solving the general case of \eqref{qncode}-\eqref{flat} which is done numerically applying a shooting method. Starting from  slightly
perturbed values for the left state, we integrate the resulting initial value
problem for a system of differential algebraic equations (DAEs), 
rewritten in terms of $\phi_s$ and
$\phi_+$ as the dependent variables. For a fixed the value of $\tilde{\mu}$, there is a unique value of $\mu_s$ such that the
trajectory connects to the right equilibrium. The critical value of $\mu_s$ is then determined using a bisection iteration. The resulting solution defines $\mu_s$ and the
associated values for the left and right state $\phi_s^a$ and $\phi_s^b$ for $\phi_s$
and similarly for $\phi_-$.
\begin{figure}[ht]
	\centering
\includegraphics[width=0.75\textwidth]{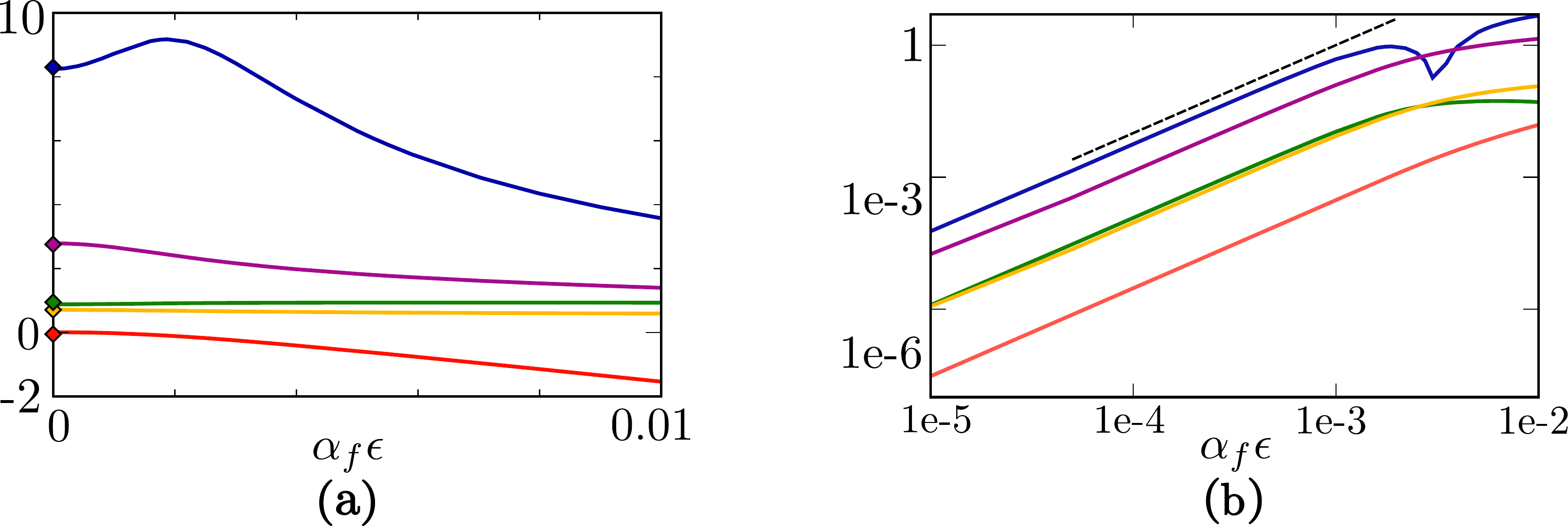}
\caption{(a) The figure shows graphs of $100\mu_s$, $\phi_s^a$, $\phi_s^b$,
$\tilde\phi_+^a$ and $\tilde\phi_+^b$, listed in order from bottom to top. 
The diamonds at $\epsilon=0$
represent the values 
 $100\mu^{\alpha}$, $\phi_s^{\alpha}$, $\phi_s^{\beta}$,
$\tilde\phi_+^{\alpha}$ and $\tilde\phi_+^{\beta}$
obtained from the asymptotic solution 
as given by \eqref{mwvals} and \eqref{phiqphip-odec}. 
(b) This log-log plot has graphs for 
$|\mu_s-\mu_s^{\alpha}|$, 
$\phi_s^a-\phi_s^{\alpha}$, $\phi_s^b-\phi_s^{\beta}$, 
$\tilde\phi_+^{a}-\tilde\phi_+^{\alpha}$, $|\tilde\phi_+^{b}-\tilde\phi_+^{\beta}|$.
The short top line represents a quadratic function, and is included to guide the eye.
}\label{asycheck}
\end{figure}
For the case of $\tilde{\mu}$ equal to zero, the results are shown in Fig.~\ref{asycheck}. The plots in (a) show graphs of
the values for $100\mu_s$ and for the values of the two equilibrium states for
$\phi_s^0$ and $\tilde\phi_+^0$ as a function of the salt concentration $\epsilon$. Note that by increasing $\ionbp$, we increase the concentration of positive ions $\phi_+$ (see Eq.~\ref{eq:dil_limit}), hence departing from the dilute limit. For $\epsilon\to 0$, the solution converges to the values given by the leading-order asymptotic
solution \eqref{maxwell}. Moreover, near to $\epsilon=0$, the
behaviour is quadratic, as can be seen from the log-log plot in
Fig.~\ref{asycheck} (b), consistent with neglecting $O(\epsilon^2)$-terms
in \eqref{phiqphip-ode}. However, for $\alpha_f\epsilon$ greater than $1\times 10^{-3}$,
the value for $\tilde\phi_{+,2}$ departs from this behaviour and in fact passes
through a maximum as it reverses its trend.
We remark that $\mu_s$ is positive
for $\alpha_f\epsilon\leq 5.45\times10^{-3}$, and negative for larger values of $\epsilon$.
For a closer comparison of the simulation with solutions from the system of
DAEs, we consider again the situation where the influx of salt and solvent is stopped
by removing the gel from the bath. In this scenario there are two additional integral constraints coming from the conservation of total solvent and ions within the gel, which can be used to fix the two remaining degrees of freedom, i.e. the location of the front and the value of $\tilde{\mu}$. For simplicity we just read out this values from the numerics and conclude the $\tilde{\mu}$ is again negligible. 
When comparing the numerical solutions with the solutions from the DAE (see red dashed line in Fig.~\ref{nofluxset1A}), these are almost in perfect agreement.
We can further apply this analysis to set 2 in~(\ref{par_sim2}) for which we move away from the dilute approximation. In Section \ref{sec:numerics_spin}, we will show that the estimates from the phase-plane analysis can be used to investigate also the structure observed in Fig.~\ref{equi1_stab} (set 2).

\section{Spinodal decomposition}
\label{sec_stability}
In the previous section we have seen how changing the properties of the bath can be sufficient to induce phase separation of the gel via development and propagation of a depletion front. In this section we are interested in another common modality of phase separation: spinodal decomposition. In this scenario, a spatially homogeneous region of the gel spontaneously separates into regions of high and low solvent content once perturbed with noise. This can occur when the homogeneous state the gel is in is unstable. As discussed in \cite{Hennessy2020}, spinodal decomposition can be induced in neutral gels by changing the temperature $T$ and hence the parameter $\chi$. While we will also discuss this scenario briefly (see Fig.~\ref{sim4}), for the polyelectrolyte gels studied here, we examine whether spinodal decomposition can also be induced by changing the ion concentration in the bath.

\subsection{Stability analysis}
As shown in Fig.~\ref{equi1_stab}, the transition between different equilibrium states can also be accompanied by phase separation via spinodal decomposition of the gel bulk (region ahead of the front). To understand this mechanism, we investigate the stability of the initial homogeneous composition of the gel bulk, here denoted by $(\bar\phi_s,\bar\phi_+)$, to small amplitude perturbations. 
As in Section~\ref{sec_phaseplane}, we rescale $z$ with $\omega$, which is the characteristic length scale of the internal interfaces; to ease the algebra we further scale $t$ with $\omega^{-1}$. Since $\omega$ is small, the homogeneous region of the gel now fills the entire space $-\infty<z<\infty$. 

We can now perturb the base state $(\bar{\phi}_s,\bar{\phi}_+)$ with normal modes suitable for an infinite domain by letting
\begin{equation}
\phi_s = \bar{\phi}_s + \delta\tilde\phi_s e^{ikz+\lambda t}, \quad 
\phi_+ = \bar{\phi}_+ + \delta\tilde\phi_+ e^{ikz+\lambda t}
\label{modes}
\end{equation}
with $k\in\mathbb{R}$, $\delta\ll 1$, and  $\lambda \in \mathbb{C}$  being the wavenumber, amplitude, and growth rate of the perturbation.
Substituting~(\ref{modes}) into the governing equations~(\ref{gov_a})-(\ref{gov_b}) and keeping only the $O(\delta)$ terms leads to the $2\times2$ system
\begin{subequations}
	\begin{align}
	\lambda  \tilde\phi_s+k^2 \left[\left(1-\eqphi{s}\right)  \tilde j_s -2\eqphi{s}  \tilde j_+ \right]=0,\\
	\lambda   \tilde\phi_+ +k^2 \left[\left(1-2\eqphi{+}\right)  \tilde j_+ -\eqphi{+}  \tilde j_s \right]=0,
	\end{align}
	where $\tilde j_s$ and $\tilde j_+$ are the solution of the linear system
	\begin{align}
	\begin{aligned}
	\left(1+ \myfrac[2pt]{\alpha_f \eqphi{n}}{\D\eqphi{s}}\right)  \tilde j_s - \myfrac[2pt]{\alpha_f\eqphi{n} }{\D\eqphi{-}}  \tilde j_+ = \left[\left(\eqphi{s}B_{\phi_+}+\eqphi{+}A_{\phi_+}\right)  \tilde \phi_+ + \right.\\
	\left.\left(k^2\eqphi{s}\left(1+\alpha_f\right)\eqphi{n}+\eqphi{s}B_{\phi_s}+\eqphi{+}A_{\phi_s}\right)  \tilde \phi_s \right]
	\end{aligned}\\[3mm]
	\left(1+ \myfrac[2pt]{\eqphi{+}}{ \eqphi{-}}\right)  \tilde j_+ - \myfrac[2pt]{2\eqphi{+}}{\eqphi{s}}  \tilde j_s =\D \eqphi{+} \left[\left(A_{\phi_+}-2k^2\eqphi{s}\right)  \tilde\phi_s +A_{\phi_+} \tilde \phi_+ \right].
	\end{align}\label{stab_elec_neu}%
\end{subequations}
The subscripts ${\phi_s,\phi_+}$ now denote the derivatives of $A$ and $B$ with respect to these variables evaluated at $(\bar{\phi}_s,\bar{\phi}_+)$. The values of $\bar{\phi}_-$ and $\bar{\phi}_n$ are defined by evaluating~(\ref{phin})-(\ref{eq:phim}) at $\bar{\phi}_s,\bar{\phi}_+$.
Note that we could have solved explicitly for $\tilde j_s$ and $\tilde j_+$ but this would not benefit the exposition given the complexity of the equations. Imposing that the system~(\ref{stab_elec_neu}) has non-trivial solution, we obtain the growth rate $\lambda=\lambda(k)$ as the roots of the characteristic polynomial:
\begin{equation}
P_k(\lambda) = \lambda^2+\frac{\zeta_{20}+ \zeta_{22} k^2}{\zeta_1} k^2\lambda +\frac{\zeta_{30}+ \zeta_{32} k^2}{\zeta_1} k^4\D ,\label{lambdaeq}
\end{equation}
where the relevant coefficients $\zeta$ are functions of $(\bar\phi_s,\bar\phi_+)$ (see supplemental material, section \ref{app_stability}) and  $\zeta_1>0$.
The two roots of~(\ref{lambdaeq}) are
\begin{subequations}
\begin{align}
    \lambda_{\pm}(k) = \frac{1}{2} \left[\T_k \pm\sqrt{ \T_k^2-4\Delta_k}\,\right],\\
    \T_k = -\frac{\zeta_{20}+ \zeta_{22} k^2}{\zeta_1}k^2,\quad
    \Delta_k=\frac{\zeta_{30}+ \zeta_{32} k^2}{\zeta_1} k^4\D.
    \end{align}\label{eq:eigenvalues}
\end{subequations}
The homogeneous state is stable if and only if the real part of $\lambda_\pm(k)$ is negative for all $k$. 
We note that $\lambda_\pm(k)$ are equivalent to the eigenvalues of the linear dynamical system $\dot{\vec y}=A_k\vec y$, where $\vec{y} \in \mathbb{R}^{2}$ and $A_k\in \mathbb{R}^{2\times 2}$ has trace $\T_k$ and determinant $\Delta_k$. We will refer to this class of dynamical systems, which is parametrised by $k$, as $\Sigma_k$. The problem of studying the stability of the mode $k$ is therefore analogous to studying the stability of $(0,0)$ for the system $\Sigma_k$. We therefore have that the $k$-th mode is stable if and only if $\T_k<0$ and $\Delta_k>0$. When considering $k\gg1$ then $\Delta_k\approx \D k^6\zeta_{32}/\zeta_{1}$ and $\T_k\approx-\zeta_{22} k^4/\zeta_1$.
Since $\zeta_{1}$, $\zeta_{22}$ and $\zeta_{32}$ are always positive the conditions of stability are always satisfied for large wavenumbers. This implies that $(\eqphi{s},\eqphi{+})$ is unstable if and only if the system $\Sigma_k$ has at least a bifurcation point $k_*>0$, where the stability of $\vec{y} = (0,0)$ changes. If $\Sigma_k$ has no bifurcation point, then $(\eqphi{s},\eqphi{+})$ is stable. Let us assume such $k_*>0$ exist then
\begin{subequations}
\begin{align}
\mbox{either} \quad \Delta_{k_*}=0 \ &\text{and} \ \T_{k_*}\leq0,\label{saddlenode}\\
\mbox{or} \quad \Delta_{k_*}\geq0 \ &\text{and} \ \T_{k_*}=0,\label{hopf}
\end{align}
\end{subequations}
where~\eqref{saddlenode} corresponds to $(0,0)$ switching from a saddle to a stable node,  while~\eqref{saddlenode} corresponds to a transition from an unstable to a stable spiral. The first scenario can occur only if $\zeta_{30}<0$ and $\zeta_{30}\zeta_{22}-\zeta_{20}\zeta_{32}\leq0$ (condition 1), while the second only if $\zeta_{20}< 0$ and $\zeta_{30}\zeta_{22}-\zeta_{20}\zeta_{32}\geq0$ (condition 2). If we denote by $\S_-$ the set of unstable homogeneous states, this is given by the union of the subsets of states satisfying condition 1 or 2. Manipulating the given inequalities, we find that the condition on $\zeta_{30}\zeta_{22}-\zeta_{20}\zeta_{32}$ does not actually play a role in determining the stability of a homogenous state and $\S_-$ is given by
\begin{equation}
    \S_-=\left\{(\eqphi{s},\eqphi{+}) \in (0,1)^2, \mbox{ s.t. }\zeta_{30}(\eqphi{s},\eqphi{+})<0 \mbox{ or } \zeta_{20}(\eqphi{s},\eqphi{+})< 0\right\}.\label{eq:Sm}
\end{equation}
The spinodal curve, which delimits the region of stability is therefore 
\begin{equation}
\begin{aligned}
    \partial\S=&\left\{(\eqphi{s},\eqphi{+}) \in (0,1)^2, \mbox{ s.t. }\zeta_{30}(\eqphi{s},\eqphi{+})=0 \:\&\: \zeta_{20}(\eqphi{s},\eqphi{+})\geq0\right\} \\&\qquad \bigcup \left\{(\eqphi{s},\eqphi{+}) \in (0,1)^2, \mbox{ s.t. }\zeta_{30}(\eqphi{s},\eqphi{+})\geq0\: \& \:\zeta_{20}(\eqphi{s},\eqphi{+})=0\right\}.
    \end{aligned}\label{eq:setspinodal}
\end{equation}

\paragraph{The limit of dilute salt concentration}
\label{sec:LSA_dilute}
Let us define $\eqphi{+}=\alpha_f \epsilon^2\tilde{\phi}_+$ (as in section~\ref{sec_phaseplane}) and take the limit $\epsilon \rightarrow 0$. 
Reasoning as in the previous section (more details in section~\ref{app_dilute} of supplemental material ), we obtain that the stability of the system is governed by the leading order approximation of the coefficients $\pert{\zeta_{20}}0$ and $\pert{\zeta_{30}}{0}$ which are of the form:
\begin{equation}
	\zeta^{(0)}_{20} =\frac{(\pert{\eqphi{n}}0)^{-1}}{1+\alpha_f}\left( \pert{B_{\phi_s}}0\eqphi{s} (1-\eqphi{s})+\left[\D+\alpha_f\eqphi{s}^{-1}\pert{\eqphi{n}}0\right]\right),\quad
	\zeta^{(0)}_{30}=\eqphi{s}\pert{B_{\phi_s}}0,
\end{equation} 
where $\pert{B_{\phi_s}}0=\pert{B_{\phi_s}}0(\eqphi{s})$ only depends on the concentration of the solvent.
It is therefore apparent that when $\pert{\zeta_{20}}0=0$ then $\zeta^{(0)}_{30}<0$, while whenever $\zeta^{(0)}_{30}=0$ then $\pert{\zeta_{20}}0>0$. Consequently, the spinoidal curve $\partial \S$ as in~(\ref{eq:setspinodal}) reduces to:
\begin{subequations}
\begin{align}
\partial\S=&\left\{\eqphi{s} \in (0,1), \mbox{ s.t. } \pert{B_{\phi_s}}0(\eqphi{s})=0\right\},
\end{align}
where
\begin{align}
\pert{B_{\phi_s}}0(\eqphi{s})=\eqphi{s}^{-1}-\frac{\left[1+2\chi(1-\eqphi{s})\right]}{1+\alpha_f}+\frac{\G}{1+\alpha_f}\left[1+\left(\myfrac[2pt]{1+\alpha_f}{1-\eqphi{s}}\right)^2\right].
\end{align}\label{spin_dilute}%
\end{subequations}
\begin{figure}[t]
	\begin{subfigure}{0.49\textwidth}
		\includegraphics[width=0.9\textwidth]{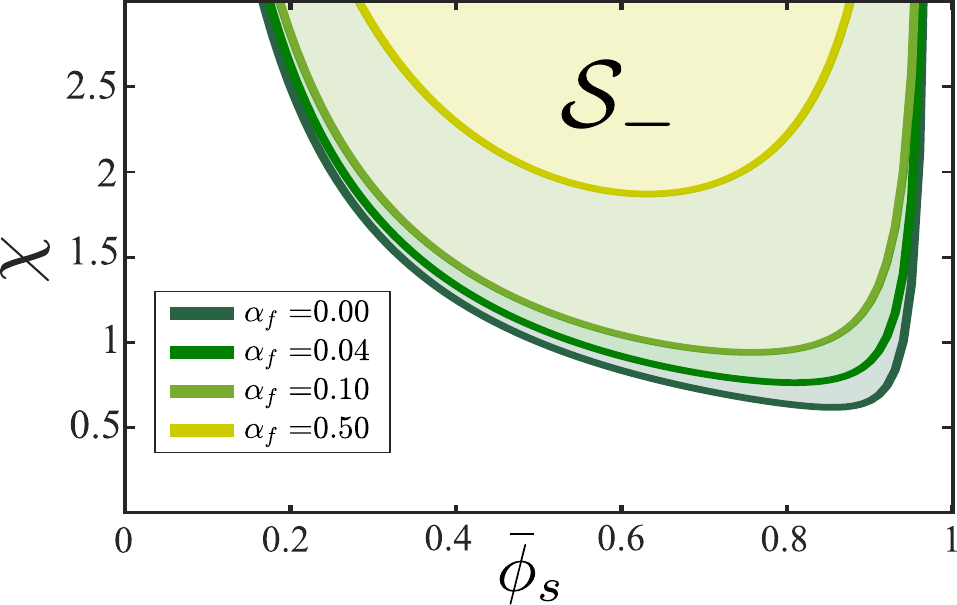}
		\caption{}
		\label{PPdilute}
	\end{subfigure}
	\begin{subfigure}{0.49\textwidth}
		\includegraphics[width=0.9\textwidth]{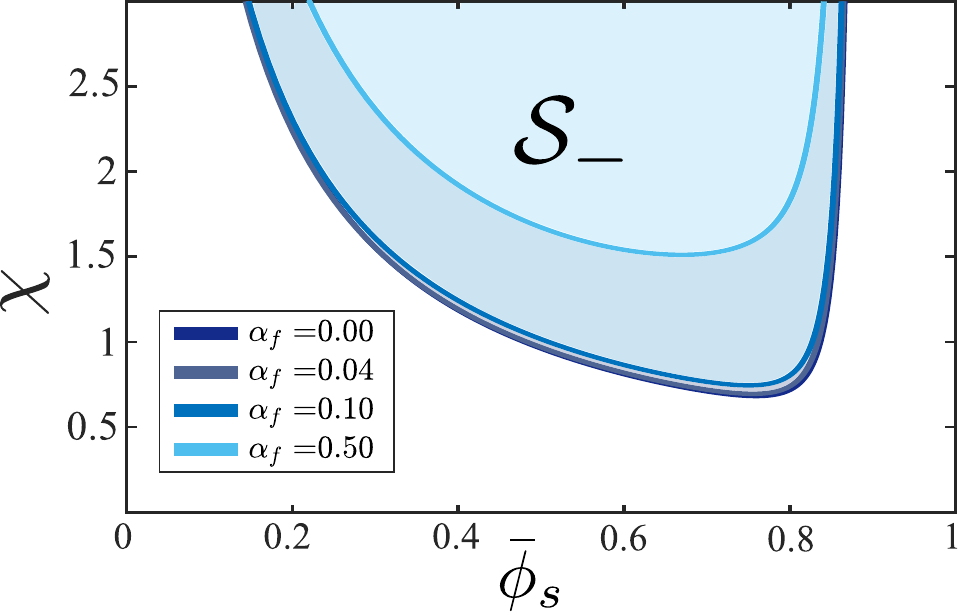}
		\caption{}
		\label{PPnodilute}
	\end{subfigure}
	\vspace{-5mm}
	\caption{Effect of the fixed charges on the stability region of an homogenous steady state for (a) dilute salt solution (b) non-dilute solution ($\bar \phi_+=0.05$). We highlight with colour the instability region $\S_-$ which is the region contained in the curve $\partial S$.}
	\label{PhasePlane1}
\end{figure}
As discussed in the supplemental material, Section~\ref{app_dilute}, Equation~(\ref{spin_dilute}) still holds for  $\alpha_f\rightarrow 0$. Furthermore, setting $\alpha_f=0$, we retrieve the same result as in \cite{Hennessy2020} for phase separation in neutral hydrogels (i.e. gel with no fixed charge on the polymer network). 
As shown in Fig.~\ref{PPdilute}, as we increase $\alpha_f$, the domain $\S_-$ shrinks which mean higher concentration of fixed charges on the polymer network can stabilise the system. Small perturbations of the homogeneous state will generate an electric field that tends to re-distribute charges homogeneously. As the number of fixed charges in the network increases so does the strength of the electric field resulting in the stabilisation of homogeneous states. As shown in Fig.~\ref{PPnodilute}, we can see a similar trend also for the non-dilute limit (more details in the following section). Fixed charges tend again to stabilise the system, however, for larger $\eqphi{+}$ the shrinking of the unstable region $\S_-$ is less than in the dilute scenario.

\paragraph{The general case}
Let us now go back to the general case of a non-dilute solutions. For the reference parameter values in~(\ref{par_sim2}), we have that $\zeta_{20}(\eqphi{s},\eqphi{+})$ is always positive. This implies that the transition to instability can only occur via a saddle-node bifurcation. Hence $\partial \S$ is implicitly defined by $\zeta_{30}(\eqphi{s},\eqphi{+})=0$ which can be computed via numerical continuation. 
\begin{figure}[t]
	\begin{subfigure}{0.32\textwidth}
		\includegraphics[width=0.95\textwidth]{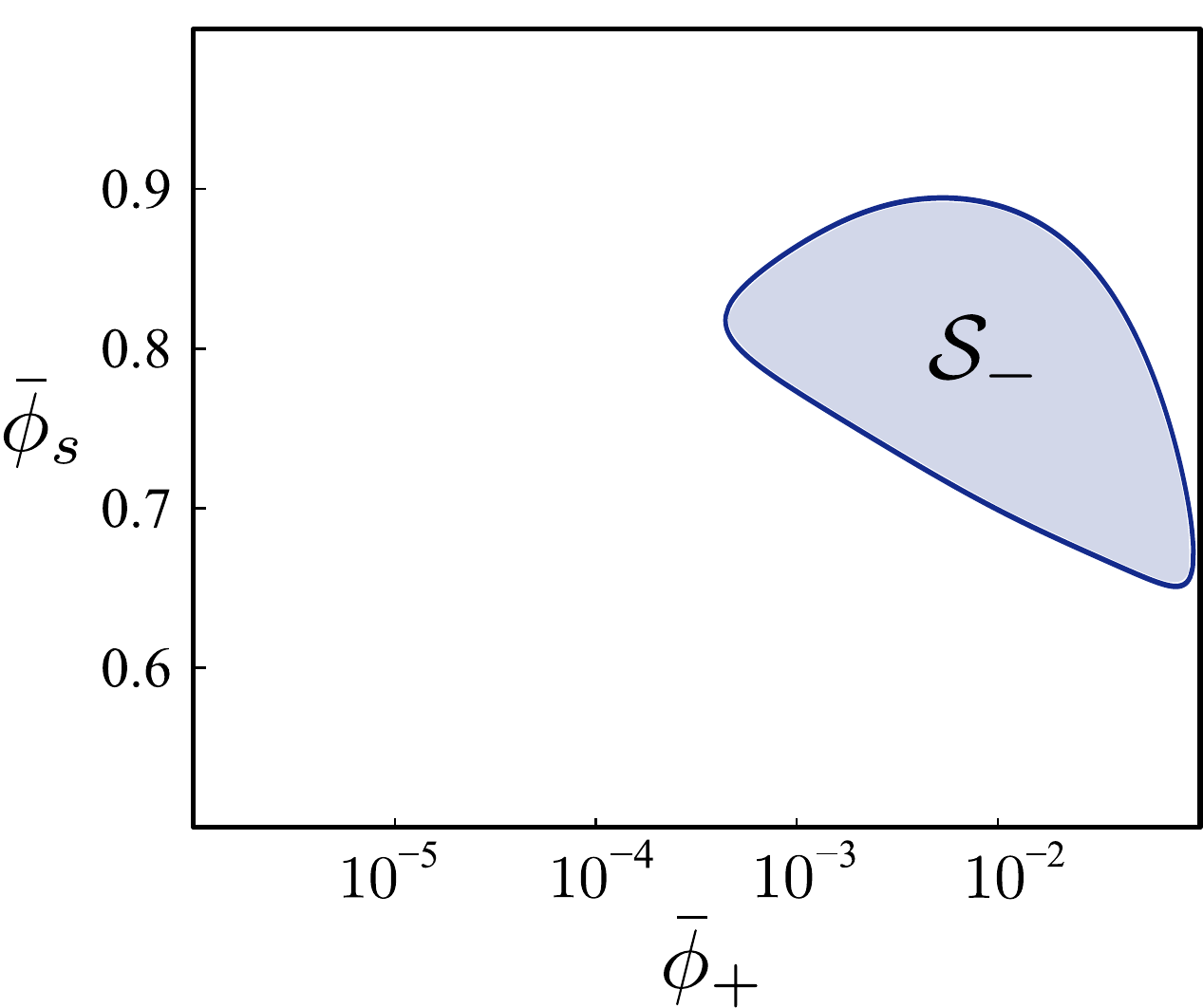}
		\caption{$\chi=0.75$}
		\label{PhasePlane2a}
	\end{subfigure}
	\begin{subfigure}{0.32\textwidth}
		\includegraphics[width=0.95\textwidth]{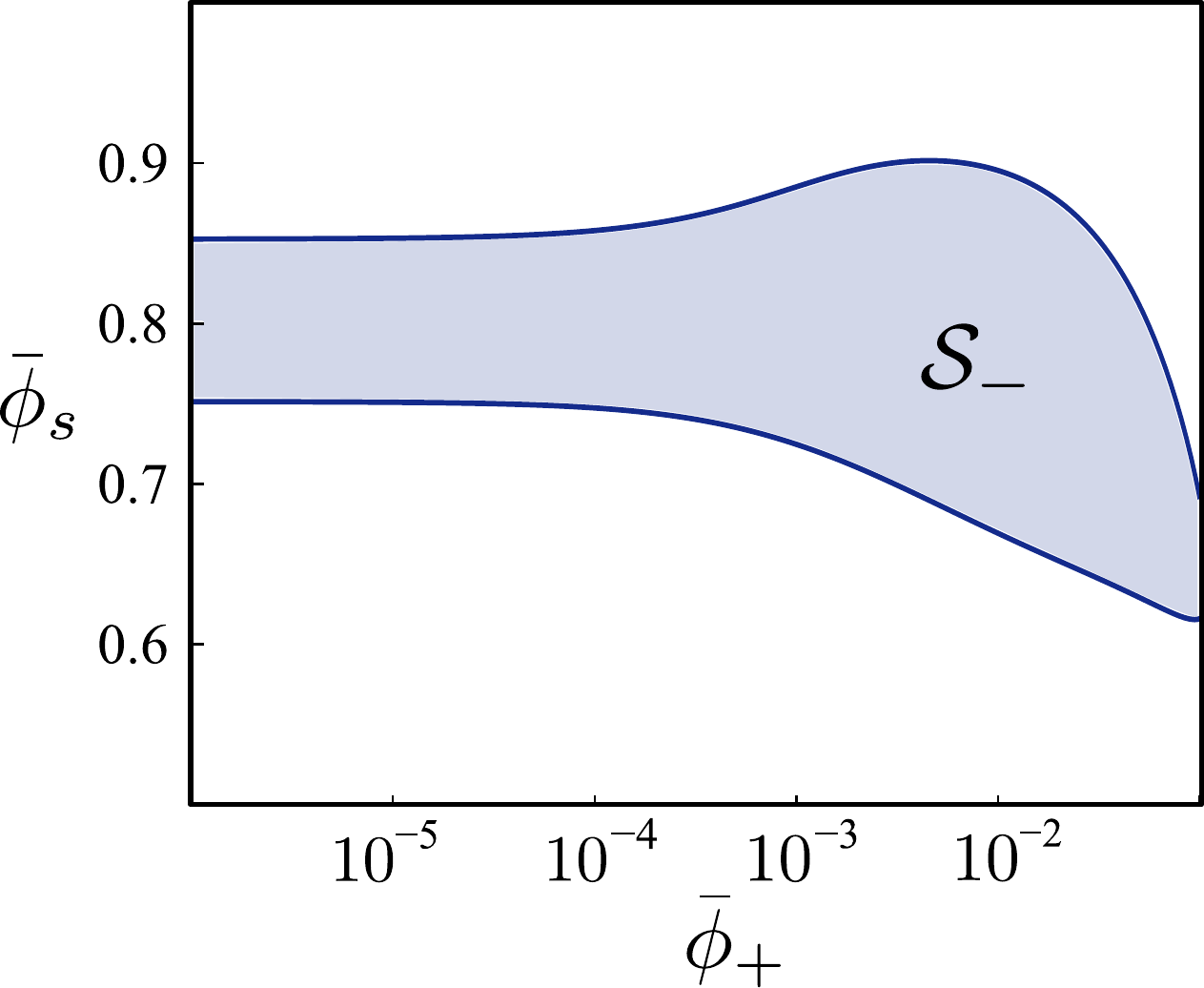}
		\caption{$\chi=0.78$}
		\label{PhasePlane2b}
	\end{subfigure}
	\begin{subfigure}{0.32\textwidth}
		\includegraphics[width=0.95\textwidth]{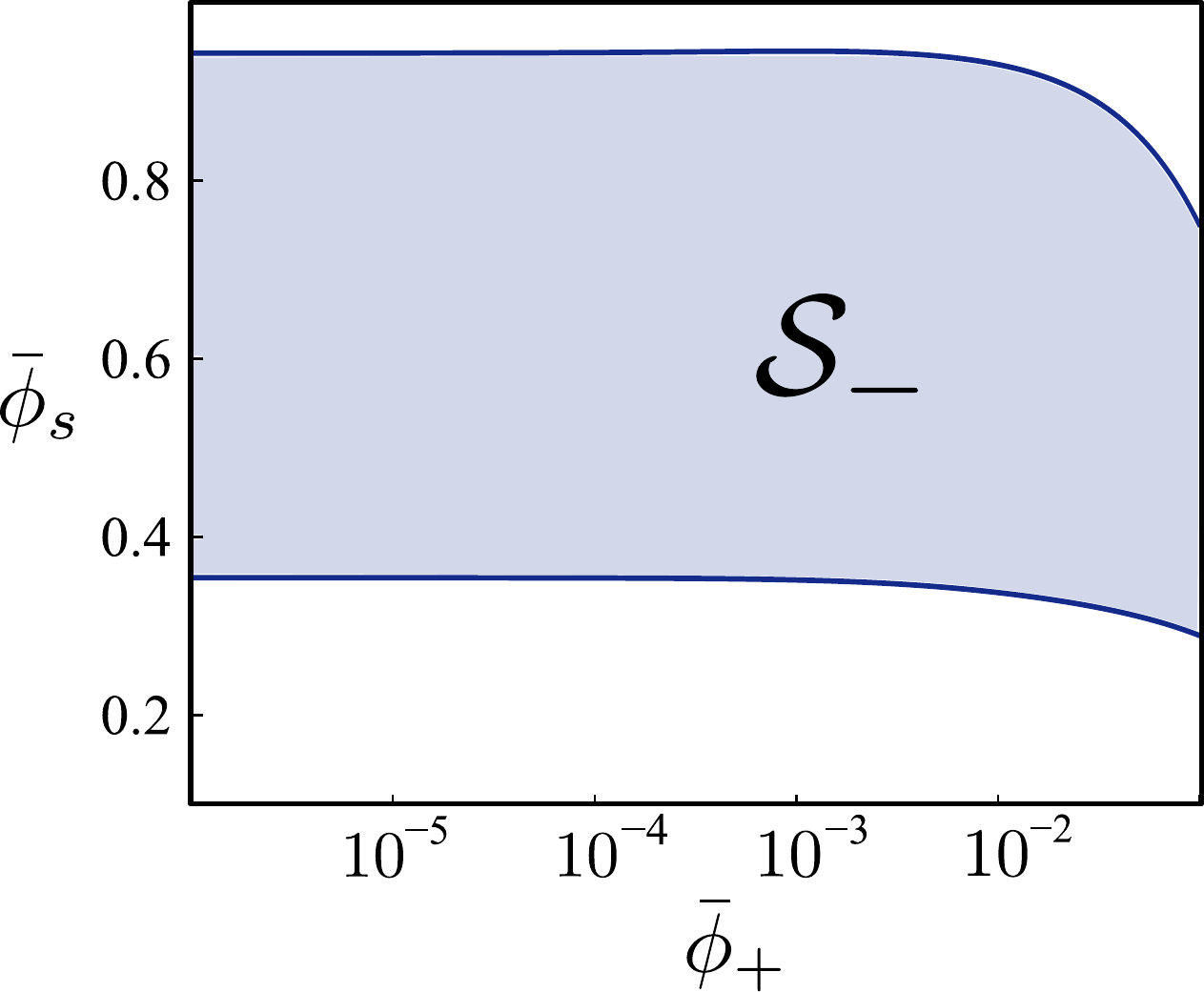}
		\caption{$\chi=1.5$}
		\label{PhasePlane2c}
	\end{subfigure}
	\vspace{-4mm}
	\caption{Plots of the instability region $\S_-$ (as defined by Eq.~(\ref{eq:Sm})) for increasing values of $\chi$.} 
	\label{PhasePlane2}
\end{figure} 
As shown in Fig.~\ref{PhasePlane2}, for concentrations $\eqphi{+}<10^{-4}$, the shape and size of the domain $\S_-$ is independent of the actual value of $\eqphi{+}$, in line with the result from the dilute section.
As we move away from the dilute limit by increasing $\eqphi{+}$, two scenarios are possible. If $\chi$ is sufficiently small (as in Fig.~\ref{PhasePlane2a}-\ref{PhasePlane2b}), then the size of $\S_-$ tends to increase with $\eqphi{+}$. This is particularly evident in Fig.~\ref{PhasePlane2a}, where in the dilute regime all homogenous states are stable, and the system only allows for spinodal decomposition to occur in the non-dilute regime, i.e. $\eqphi{+}\approx 10^{-3}$. If we are however to increase $\eqphi{+}$ further (i.e. $\eqphi{+}\approx 10^{-2}$) then $\S_-$ starts to shrink and the unstable state corresponds to a less swollen gel. On the other hand, if $\chi$ is larger (such as Fig.~\ref{PhasePlane2c}), then increasing $\eqphi{+}$ only results in the shrinking of the instability region.

Based on stability analysis, we now want to identify how we can drive spinodal decomposition in a gel. A standard approach is to increase the value of $\chi$ (i.e the temperature) \cite{Hennessy2020}. This exploits the fact that size of the domain $\S_-$ increases with $\chi$ and moving along a vertical line in Fig.~\ref{PhasePlane1} can push the system into the unstable regime. 
\begin{figure}[t]
	\begin{subfigure}{0.5\textwidth}
		\includegraphics*[width=\textwidth]{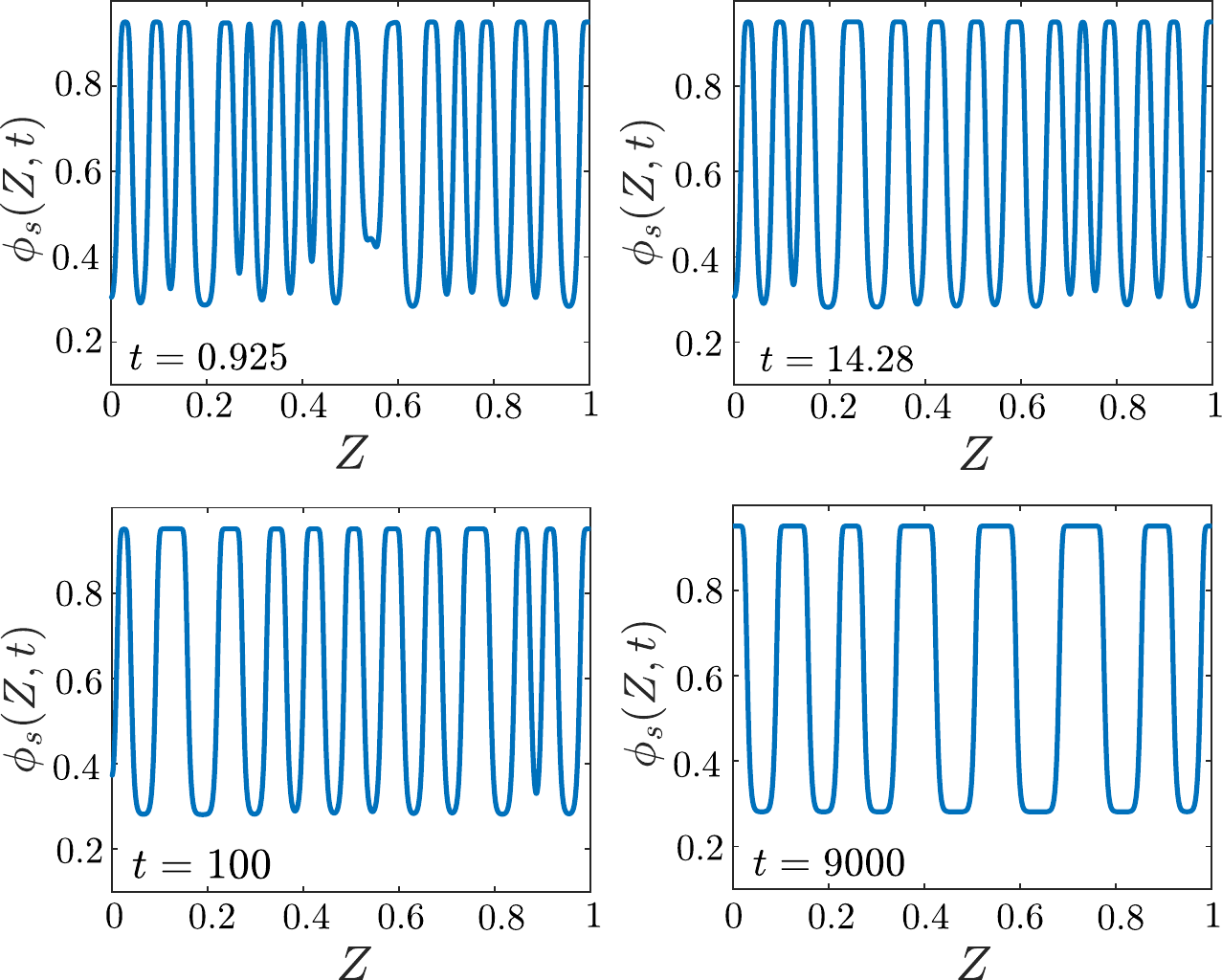}
	\end{subfigure}
	\hspace{5mm}
	\begin{subfigure}{0.45\textwidth}
		\includegraphics*[width=0.95\textwidth]{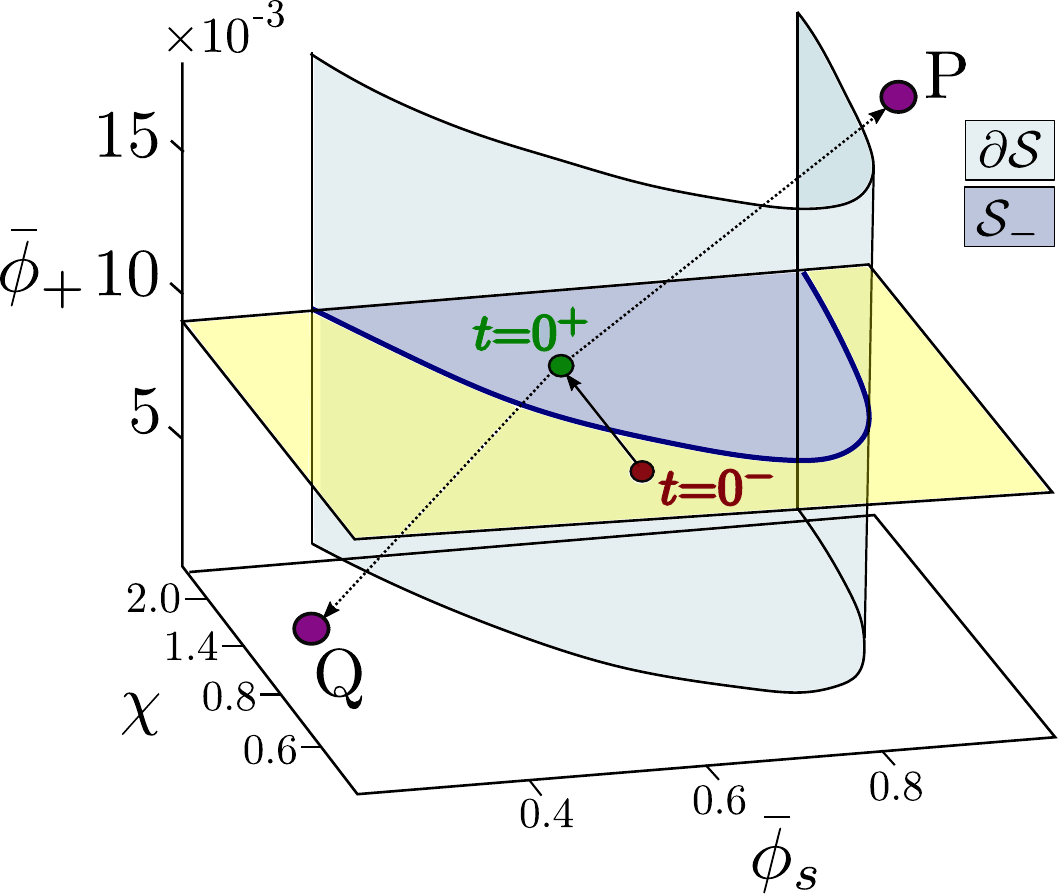}
	\end{subfigure}
	\caption{Spinodal decomposition in an isolated gel. Initially (see red point in the phase plane) the gel is equilibrated with a bath ($\phi_{0^-}=5\times10^{-3}$). At time $t=0$, we isolate the gel,
		increase $\chi$ from $0.78$ to $ 1.2$ (see green point in the phase plane) and introduce some noise. On the left, we illustrate the evolution of the $\phi_s$; on the right, we identify the manifold $\partial \S$ that divides the stable (yellow) and unstable (blue) region. The point $P$ and $Q$ represent the peak and the trough of the spikes. We use the values  $\G=0.001$, $\omega=0.01$, $\alpha_f=0.04$ and $\D=5$. }
	\label{sim4}
\end{figure}
We explore this scenario numerically, by consider an initially homogeneous gel in equilibrium with an ionic bath (parameter values $\phi_0=5\times 10^{-3}$ and
$\chi=0.78$). 
At time $t=0$, the gel is isolated and the temperature is raised so that $\chi$ increases to $\chi=1.2$. As shown in the phase plane in Fig.~\ref{sim4} this is sufficient to move the system from the stable region (see the red point in the phase plane of Fig.~\ref{sim4}) into the unstable
regime (see the green point in the phase plane of Fig.~\ref{sim4}). As we introduced some noise in the system, growing perturbations rapidly fill the entire length of the
gel and then begin to coarsen or collide, resulting in fewer and broader spikes. As shown in the phase plane in Fig.~\ref{sim4}, the peak (point P) and trough (point Q) of the spikes are located in the region of linear stability. Hence, no further instabilities develop. 
Over time, the evolution slows down until the pattern is almost stationary. However,
we expect that in principle coarsening continues until only two regions remain,
one in the collapsed state with $\phi_s \approx 0.3$ and the other in the swollen state with $\phi_s \approx 0.9$. These two end-state values are stable as indicated in the right-most panel in the figure. 

\subsection{Spinodal decomposition of a collapsing gel}
\label{sec:numerics_spin}
The stability analysis however hints at another possible mechanism to drive spinodal decomposition in polyelectrolyte gels: increasing the concentration of co-ions in the system. Experimentally, this can be achieved maintaining the gel in contact with the bath and increasing $\phi_0$. As discussed in Section~\ref{sec_collapse}, the ions rapidly diffuse in the gel bulk while the solvent concentration remains constant in this fast transient. Exploiting the different time scales in the system, we can therefore move along horizontal lines in the phase plane of Fig.~\ref{PhasePlane2}. For $\chi=0.78$ (as in the simulation in Section~\ref{sec_collapse}), changes in the instability region $\S_-$ occur only when $\phi_+\sim O(10^{-3})$. Using the equation~(\ref{estphipm}), we obtain that $\ionbp$ must be increased to $\ionbp\sim O(10^{-2})$ in order to exploit the growth of $\S_-$. 
\label{sec:spin_bath}
\begin{figure}[t]
	\includegraphics[width=\textwidth]{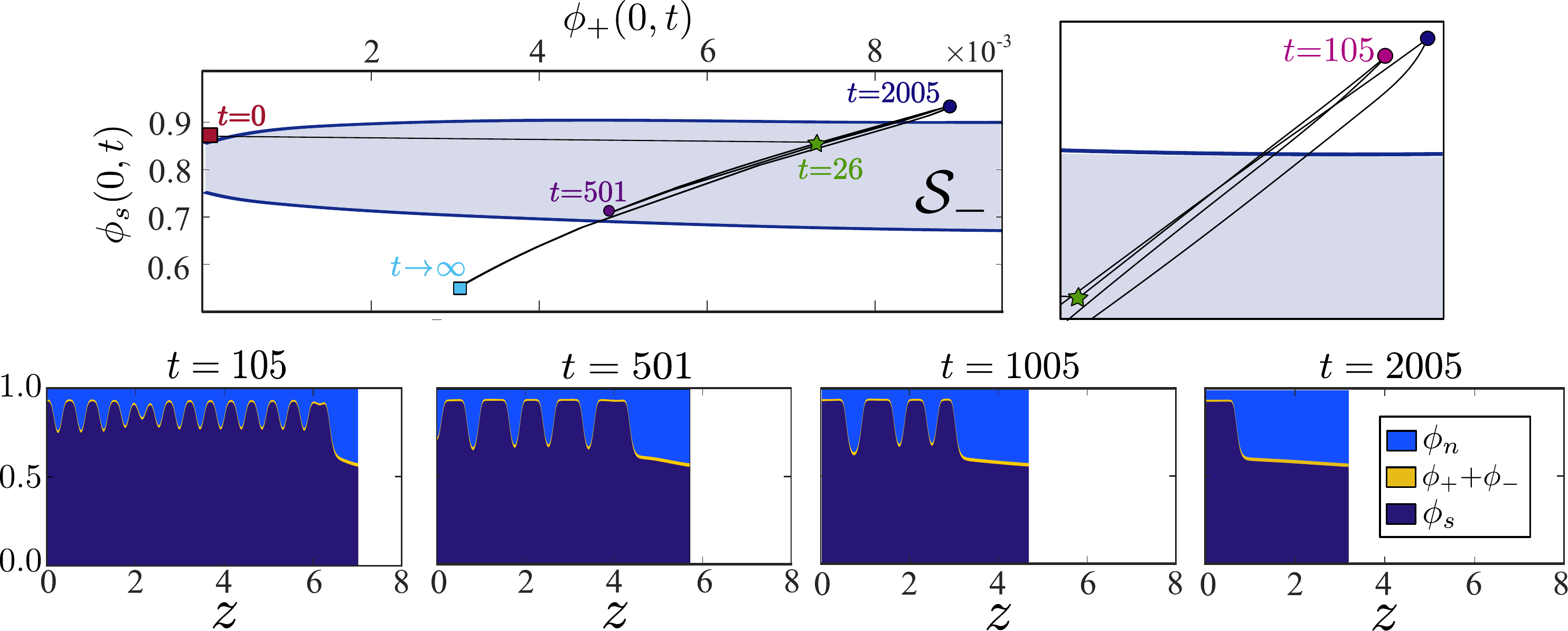}
	\caption{Spinodal decomposition in a collapsing gel: (top) time evolution of the gel composition $(\phi_s,\phi_+)$ at location $z=0$. The instability region $\S_-$ as predicted by the stability analysis is highlighted in blue; the panel on the right is a zoom near the green star and the blue dot to better visualise the complexity of the trajectories; (bottom) snapshot of the gel composition at different time points. Parameter values corresponding to set 2 in (\ref{par_sim2}).}
	\label{set3A}
\end{figure}

\begin{figure}[t]
	\centering
	\includegraphics[width=0.95\textwidth]{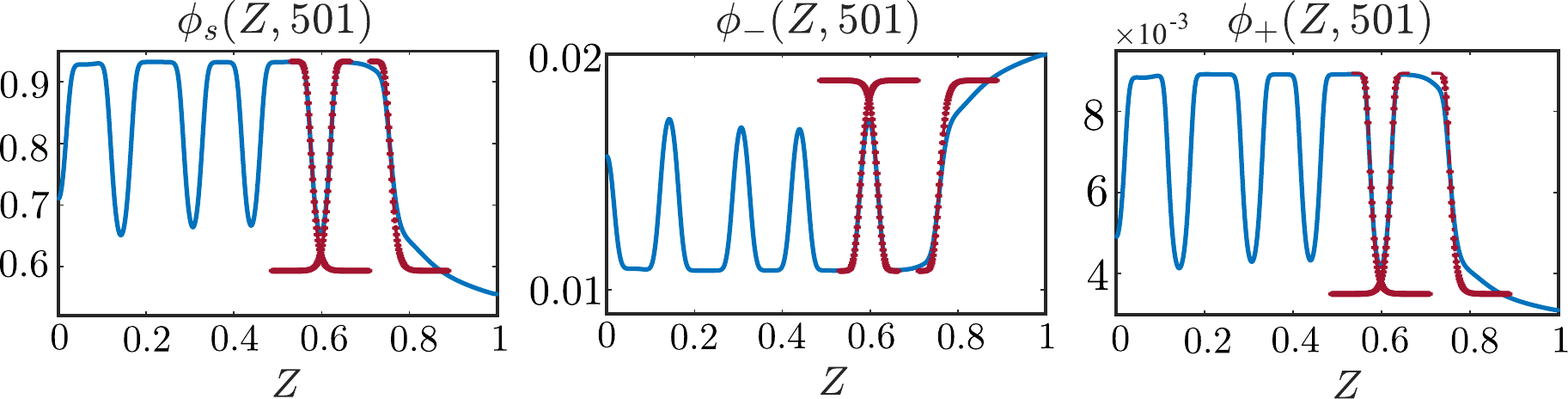}
	\caption{Comparison of the phase plane solution for $\phi_s$, $\phi_-$ and $\phi_+$
		(shown with dashed red lines and symbols) with the numerical results
		at time $t=501$ in
		Fig.~\ref{set3A}.}\label{set2comp_phaseplane}
\end{figure}

An example of this scenario is shown in Fig.~\ref{set3A}, where we present the results of the numerical
simulation for set 2 in~(\ref{par_sim2}). As expected the ions rapidly diffuse in the gel, driving the bulk of the gel into the unstable region of the phase diagram; see $t=26$ in the right panel of Fig.~\ref{set3A}. This results in the onset of spinodal decomposition, which give rise to a series of solvent-depleted phases, which coarsen and collide first rapidly and then very
slowly. Interestingly, at the centre of each region with a low solvent concentration there is also a high counter-ion ($\phi_-$) concentration, as a consequence of maintaining electro-neutrality (see Fig.~\ref{set2comp_phaseplane}). As shown in Fig.~\ref{set3A}, on a longer time scale, the depletion front propagates into the gel, consuming the array of solvent-depleted domains. When comparing with the result in Fig.~\ref{set1A}, we note that the front propagates faster in this case, with spinodal decomposition facilitating the removal of solvent from the gel. 

Again, we can use the phase-plane analysis (see Sec. \ref{sec_phaseplane}) to predict the structure of the collapsing gel. When considering the depletion front, the full model and the phase-space analysis prediction deviate slightly. As shown in Fig.~\ref{set2comp_phaseplane}, the front obtained via the phase-plane analysis is too steep and its lower limit does not capture the actual right state of the front very well. The interface of the solvent-depleted sub-domains forming in the bulk of the gel yields a better comparison. A possible explanation of the higher discrepancy is that the front evolves more rapidly so that the quasi-stationary assumption underlying the phase plane analysis breaks down. This is located near the free interface, where we have larger variations in the chemical potentials and therefore fluxes so that the quasi-stationary assumption underlying the phase-plane analysis may not hold.

\section{Conclusions}\label{sec:conclusions}
In this study we have focussed on the transient dynamics of a
polyelectrolyte gel in contact with a ionic bath containing a monovalent salt.
Starting from a homogeneous equilibrium, we initiate the collapse of the gel by changing the surrounding salt concentration. Depending
on the properties of the system, temporary patterns form with highly swollen
and collapsed regions co-existing in the gel. Using the mean-field model
developed in the companion paper \cite{Celora_modelling_2020}, we can track the
evolution of these internal interfaces, delineating the regions where the gel
network has collapsed after they have formed via phase separation.
We here consider the case of a one-dimensional constrained geometry, with a
single free interface in contact with the bath. The model is further simplified
by passing to the electroneutral limit, justified by the infinitesimal size of
the electric double layer compared to the other spatial scales in the model. 

Depending on the value of various model parameters, such as the Flory
interaction parameter $\chi$, the gel shear modulus $\G$, and the ratio of the
salt concentration in the bath and the fixed charges on the polymer network
$\epsilon=\ionbp/\alpha_f$, the stability properties of the homogeneous states vary.
Our numerical results for the case where the gel is isolated from the bath (by
using no-flux boundary conditions at the free interface) show that as the unstable
homogeneous state undergoes spinodal decomposition, it forms arrays of locally
collapsed solvent-depleted domains with a high concentrations of counter-ions. These
eventually coarsen into a new stable equilibrium. 

When in contact with the bath, the gel is sensitive to the salt concentration
$\ionbp$ in the surrounding fluid. Increasing $\ionbp$ initiates and drives the
collapse of the gel. This process is  characterised by a depletion front that
forms via phase separation and travels into the gel.  Using phase-plane
analysis, we are able to show that the depletion front selects the homogeneous
states at its front and rear. These are always linearly stable. In the
spinodally unstable regime, the emerging structure of collapsed
solvent-depleted domains are consumed by the depletion front that propagates
through the gel on a faster time scale than the coarsening process.  

Even though our analysis is currently only one-dimensional, it sheds light on
the possibility of observing these patterns experimentally. Depending on
parameter settings, the transient patterns with high concentration of $\phi_-$
in the collapsed and high concentration of $\phi_+$ in the swollen regions
are in principle observable. 
Our approach provides a template for how mathematical techniques such as the phase-plane analysis and stability analysis can elucidate the notoriously complicated models
for polyelectrolytes, beyond numerical simulations, particularly when employed in combination. 

An important consequence of our approach is that fundamental quantities, such as the velocity of the localised front, can now be predicted using an asymptotic analysis based on our approximation of the depletion front.
Our analysis can also be easily genralised to settings that account for example for multivalent salts and concentration-dependent permittivity and may allow to capture further 
scenarios of collapse. This will be carried out in our forthcoming work, in the one-dimensional setting as well as in higher-dimensional formulations. 

\clearpage
\appendix

\section{Table of parameters}

\mbox{}\relax

\begin{table}[h]
\caption{Physical parameters used in the full dimensional model 
\eqref{model3D1}, \eqref{constlaws}. Unless indicated otherwise, values are taken from
\cite{Yu2017}, except for universal/generic constants that can
be obtained from standard sources.
}\label{tab:physparms}
\begin{center}
\begin{tabular}{lp{0.35\textwidth} @{\hskip 0.5in} p{0.375\textwidth}}
\toprule[1.5pt]\addlinespace[2pt]
& \textbf{Meaning} & \textbf{Typical value(s)}  \\ \hline\addlinespace[2pt]
${\D}_\pm$ & Diffusivity of mobile ions in solvent. 
& $\D_\pm=\D \approx 10^{-9}$ m$^2$ s$^{-1}$  \cite{SherwoodThomasK1975Mt}
\\ \hline\addlinespace[1.5pt]
${\D}_s$ &Solvent diffusivity in gel& ${\D}_s\approx 10^{-11}\ldots 10^{-9}$ m$^2$ s$^{-1}$ \cite{Drozdov2016b}
\\\hline\addlinespace[2pt]
$k_B$ & Boltzmann constant & $1.38\times 10^{-23}$ JK$^{-1}$  
\\ \hline\addlinespace[2pt]
$T$ & Temperature & $300$K   \\ \hline\addlinespace[2pt]
$\chi$ &Flory-Huggins Parameter&$0.1,\ldots, 2.3$ \\\hline\addlinespace[2pt]
$\nu$  &Volume per molecule of mobile species and fixed charges &  $10^{-28}$ m$^3$  \\\hline\addlinespace[2pt]
$\alpha_f$ & volume fraction of fixed charge per unit volume of dry gel & $\alpha_f\approx0.01,\ldots, 0.1$
 \\\hline\addlinespace[2pt]
$z_f$ & valence of the fixed charges &  here $z_f=1$
 \\\hline\addlinespace[2pt]
 $z_\pm$ & valence of the mobile ions &  here $z_\pm=\pm 1$
 \\\hline\addlinespace[2pt]
$G$ &shear modulus& $k_B TN_p$
\\\hline\addlinespace[2pt]
$N_p$ & Number density of polymer chains in the dry state & $N_p \nu\approx10^{-4}\ldots10^{-3}$
\\\hline\addlinespace[2pt]
$\phi_0$ &Far-field salt fraction in bath& $\phi_0\approx10^{-7}\ldots 10^{-1}$ 
\\\hline\addlinespace[2pt]
$L$ & Macroscopic dimension of gel in the unconstrained direction & $0.001\ldots 0.01$ m
 \\\hline\addlinespace[2pt]
$\gamma$ & Interface stiffness parameter & Assumed to satisfy
$ L_{\text{int}}\ll L$ in the main text.\\\hline\addlinespace[2pt]
$K$ & Permeability of solvent in gel,
$K={\D_s \phi_n^{-\theta}}/{k_BT}$ & $\theta\geq0$ \cite{Drozdov2016b}\newline we here set $\theta$=0 in the simulation as in \cite{Hennessy2020}
\\\bottomrule[2pt]
\end{tabular}
\end{center}
\end{table}

\section{Numerical methods}
\label{numerics}
For the numerical implementation of the electro-\-neut\-ral model, it is convenient to resize the moving domain $[0,h(t)]$ onto a fixed one using the following change of variables:
\begin{equation}
Z= \frac{z}{h(t)}, \quad \frac{d h}{dt}=\left.v_n\right|_{h(t)}.
\end{equation}

At each time point $t_i$ we solve for $\vec{y}=(\phi_s,\phi_+,j_s,j_+)$. We discretise the governing equation~(\ref{gov_a})-(\ref{gov_b}) using finite difference method on a staggered-grid grid for the spatial dimension $Z$ and a semi-implicit method as a time-stepping method.

\subsection*{Isolated gel} For the case of an isolated gel, the fluxes are evaluated on the edges while volume fractions and chemical potentials on the cell midpoints. In this case we treat linear term implicitly while approximate non-linear term explicitly \cite{Hennessy2020}.
For the boundary condition, we strongly impose the no-flux boundary conditions, while the Neumann condition of $\phi_s$ is approximated using ghost point. 

\subsection*{Gel in contact with the bath} In this scenario, given the different boundary conditions, we evaluate fluxes on the cell midpoints, while the volume fractions and chemical potentials on the cell edges. Again for the time evolution we use a semi-implicit method that treats linear term in~(\ref{gov_a})-(\ref{gov_b}) implicitly while approximate non-linear term explicitly \cite{Hennessy2020}. However we keep the the non-linearity in the boundary conditions \eqref{muscont}-\eqref{mupmcont}.
The no-flux boundary conditions are strongly imposed (by adding $Z=0$ to grid for the fluxes), while the Neumann condition of $\phi_s$ is approximated using ghost point. Less trivial instead is the treatment of the non-linear terms in~(\ref{muscont})-(\ref{mupmcont}), which relies on fixed point iteration. Consider we are solving for the time $t=t_{i+1}$ and we have the solution $\vec{y}^i$ at the previous time point $t_i=t_{i+1}-dt$. Then we denote by $\vec{y}^{i+1}_j$ the $j$-th iteration of the fixed point method. We therefore treat the left-hand side of~(\ref{muscont})-(\ref{mupmcont}) implicitly, while we evaluate the right-hand side at the attempt solution $\vec{y}^{i+1}_j$. We can therefore discretise the problem, which reduces to the solving the linear system:
\begin{equation}
A(\vec{y}^i)\vec{y}^{i+1}_{j+1}= \vec{b}(\vec{y}^i,\vec{y}^i_j),
\end{equation} 
where the dependency of the vector $\vec{b}$ on $\vec{y}^i$ comes from the governing equation, while the dependency on $\vec{y}^i_{j}$ from the boundary condition. We therefore iterate over $j$ until the difference between two iteration is less than the tolerance $\|\vec{y}^i_{j+1}-\vec{y}^i_{j}\|<toll$. If the number of step required to match the tolerance is above a set limit $N_{max}$ then the time step $dt$ is decreased. We also add an extra check on the solution to make sure that the volume fraction $\phi_s$, $\phi_+$ and $\phi_n$ (which can be evaluated from the other two) have values between $0$ and $1$. If this is not the case, the time step $dt$ is decreased.

\section{Stability analysis of the homogeneous states}
\label{app_stability}
In this section we present in more details the mathematical steps required to derive Equations~(\ref{stab_elec_neu}) in Section \ref{sec_stability}.
Recalling that we use the following ansatz for the form of the normal modes: 
\begin{equation}
\begin{aligned}
\phi_m = \bar{\phi}_m +\delta \tilde\phi_m e^{ikz+\lambda t},
\end{aligned}
\end{equation}
we can substitute the above in the electro-neutral formulation of the model~(\ref{gov_a})-(\ref{gov_b}) to obtain:

\begin{subequations}
\begin{align}
\lambda  \tilde \phi_s+k^2 \left[\left(1-\eqphi{s}\right) \tilde j_s -2\eqphi{s}  \tilde j_+ \right]=0,\\
\lambda  \tilde \phi_+ +k^2 \left[\left(1-2\eqphi{+}\right) \tilde j_+ -\eqphi{+} \tilde j_s \right]=0,\\
\begin{aligned}
\left(1+ \myfrac[2pt]{\alpha_f \eqphi{n}}{\D\eqphi{s}}\right) \tilde j_s - \myfrac[2pt]{\alpha_f\eqphi{n} }{\D\eqphi{-}}  \tilde j_+ = \left[\left(\eqphi{s}a_{s+}+\eqphi{+}a_{++}\right) \tilde \phi_+ \right.\\
\left.\left(k^2\eqphi{s}\left(1+\alpha_f\right)\eqphi{n}+\eqphi{s}a_{ss}+\eqphi{+}a_{+s}\right) \tilde \phi_s \right]
\end{aligned}\\[3mm]
\left(1+ \myfrac[2pt]{\eqphi{+}}{ \eqphi{-}}\right) \tilde j_+ - \myfrac[2pt]{2\eqphi{+}}{\eqphi{s}} \tilde j_s =\D \eqphi{+} \left[\left(a_{+s}-2k^2\eqphi{s}\right) \tilde \phi_s +a_{++} \tilde \phi_+ \right].
\end{align}
\end{subequations}

In this case the explicit forms of the coefficients $a_{ij}$:
\begin{subequations}
	\begin{align}
B_{\phi_s}= \frac{\G\eqphi{n}^{-2}(\eqphi{n}^2+1)-1+\chi(\eqphi{s}-1)}{1+\alpha_f}+\eqphi{s}^{-1}-\chi\eqphi{n},\\
B_{\phi_+}=2\, \frac{\G\eqphi{n}^{-2}(\eqphi{n}^2+1)-1+\chi(\eqphi{s}-1)}{1+\alpha_f},\\
A_{\phi_s}= \frac{2}{1+\alpha_f}\left(\G\eqphi{n}^{-2}(\eqphi{n}^2+1)+\chi \eqphi{s}-1-\frac{\alpha_f\eqphi{-}^{-1}}{2}\right)-2\chi\eqphi{n},\\
A_{\phi_+}= \frac{4}{1+\alpha_f}\left(\G\eqphi{n}^{-2}(\eqphi{n}^2+1)+\chi \eqphi{s}-1+\frac{1-\alpha_f}{4}\eqphi{-}^{-1}\right)+\eqphi{+}^{-1}.
\end{align}\label{coeffa}
\end{subequations}
Imposing that the system to have non-trivial solution, we obtain a quadratic equation that defined the modes $\lambda$:
\begin{equation}
\zeta_1 \lambda^2+(\zeta_{20} + \zeta_{22} k^2)k^2\lambda +(\zeta_{30} + \zeta_{32} k^2) k^4\D =0,
\end{equation}
where the relevant coefficients $\zeta$ are defined as:
\begin{subequations}
\begin{align}
\zeta_1=\myfrac[2pt]{\eqphi{s}\eqphi{n}^{-1}\D\left(\eqphi{+}+\alpha_f\eqphi{n}\right)+\alpha^2_f\phi_n}{\D \eqphi{s}(\eqphi{+}+\alpha_f\eqphi{n})\left(1+\alpha_f\right)},\label{alpha1}\\
\begin{aligned}\label{alpha2}
\zeta_{20} =
\frac{ \eqphi{s}B_{\phi_s}+\eqphi{+}B_{\phi_+}}{1+\alpha_f} \myfrac[2pt]{\eqphi{+}+\eqphi{-}+\alpha^2_f\eqphi{n}}{\eqphi{+}+\alpha_f \eqphi{n}}+\D\eqphi{+}\eqphi{n}^{-1}\frac{(1-2\eqphi{+})A_{\phi_+}-2\eqphi{s}A_{\phi_s}}{1+\alpha_f}\\
+\frac{\eqphi{+}\eqphi{s}^{-1}\eqphi{n}^{-1}}{1+\alpha_f}\left[\alpha_f \eqphi{s}B_{\phi_+}+2\eqphi{s}\eqphi{-}A_{\phi_s}+\eqphi{-}(2\eqphi{+}+\alpha_f)A_{\phi_+}\right],
\end{aligned}\\[2mm]
\zeta_{22}=\frac{\eqphi{s}\eqphi{-}^{-1}\eqphi{n}^{-1} }{1+\alpha_f} \left(2\eqphi{+}\eqphi{n}^2\left(1+\alpha_f^2\right)+\eqphi{n}^3\alpha_f\left(1+\alpha_f\right)^2+4\D\eqphi{s}\eqphi{+}\eqphi{-}\right),\\[2pt]
\begin{aligned}\label{alpha3}
\zeta_{30}=\eqphi{+}\eqphi{s}(A_{\phi_+}B_{\phi_s}-B_{\phi_+}A_{\phi_s}),\\[2pt]
\zeta_{32}=\eqphi{+}\eqphi{s}(A_{\phi_+}\left(1-\eqphi{s}\right)+2B_{\phi_+}\eqphi{s}),
\end{aligned}
\end{align}%
\end{subequations}
we note that $\zeta_1$ and $\zeta_{22}$ are independent of $k$ and always positive. Expanding the coefficient $\zeta_{32}$, we can re-write this as:
\begin{equation}
\zeta_{32}=\omega^2\left[\frac{4\G(1+\eqphi{n}^{-2})}{1+\alpha_f}+(1+\alpha_f)\eqphi{n}\eqphi{+}^{-1}+\frac{(1-\alpha_f)^2}{1+\alpha_f}\eqphi{n}\eqphi{-}^{-1}\right],
\end{equation}
which is always positive as well.
As discussed in Section \ref{sec_stability}, the system is stable provided that $\zeta_{30}>0$ and $\zeta_{20}>0$ which corresponds to :
\begin{subequations}
\begin{align}
A_{\phi_+}B_{\phi_s}-B_{\phi_+}A_{\phi_s}>0\label{elec_neu_stab_conda}\\
\begin{aligned}
\frac{ \eqphi{s}B_{\phi_s}+\eqphi{+}B_{\phi_+}}{1+\alpha_f} \myfrac[2pt]{\eqphi{+}+\eqphi{-}+\alpha^2_f\eqphi{n}}{\eqphi{+}+\alpha_f \eqphi{n}}+\D\eqphi{+}\eqphi{n}^{-1}\frac{(1-2\eqphi{s})A_{\phi_+}-2\eqphi{s}A_{\phi_s}}{1+\alpha_f}\\
+\frac{\eqphi{+}\eqphi{s}^{-1}\eqphi{n}^{-1}}{1+\alpha_f}\left[\alpha_f \eqphi{s}B_{\phi_+}+2\eqphi{s}\eqphi{-}A_{\phi_s}+\eqphi{-}(2\eqphi{+}+\alpha_f)A_{\phi_+}\right]>0,
\end{aligned}
\end{align}\label{elec_neu_stab_cond}
\end{subequations}
\subsection{Dilute limit}
\label{app_dilute}
When considering the limit of small volume fraction of the free ions, i.e. $\eqphi{+}=\alpha_f\gamma^2 \tilde{\phi}_+$, where $\tilde{\phi}_+ \sim O(1)$ and $\epsilon=\phi_0/\alpha_f \rightarrow 0$, we can expand~(\ref{coeffa}) as an increasing sum of power of $\epsilon$, we obtain:
\begin{subequations}
	\begin{gather}
	B_{\phi_s}= \G\myfrac[2pt]{(\eqphi{n}^{(0)})^{2}+1}{\eqphi{n}^{(0)}\left(1-\eqphi{s}\right)}-\frac{1}{1+\alpha_f}-2\chi\eqphi{n}^{(0)}+\eqphi{s}^{-1}+O(\epsilon^2),\\
	B_{\phi_+}= 2\, \G\myfrac[2pt]{(\eqphi{n}^{(0)})^{2}+1}{\eqphi{n}^{(0)}\left(1-\eqphi{s}\right)}-\frac{2}{1+\alpha_f}-2\chi\eqphi{n}^{(0)}+O(\epsilon^2)\\
	A_{\phi_s}= 2\, \G\myfrac[2pt]{(\eqphi{n}^{(0)})^{2}+1}{\eqphi{n}^{(0)}\left(1-\eqphi{s}\right)}+\frac{2\chi(2\eqphi{s}-1)}{1+\alpha_f}-\frac{2}{1+\alpha_f}-\myfrac[2pt]{1}{1-\eqphi{s}}+O(\epsilon^2),\\
	A_{\phi_+}=\frac{1}{\epsilon^2\alpha_f\tilde{\phi}_+} + 4\, \G\myfrac[2pt]{(\eqphi{n}^{(0)})^{2}+1}{\eqphi{n}^{(0)}\left(1-\eqphi{s}\right)}\frac{4\left(\chi \eqphi{s}-1\right)}{1+\alpha_f}+\myfrac[2pt]{(1-\alpha_f)}{\alpha_f(1-\eqphi{s})}+O(\epsilon^2),
	\end{gather}
\end{subequations}
where $\eqphi{n}^{(0)}$ is determined by expanding~(\ref{phin}):
\begin{equation}
\pert{\eqphi{n}}0= \frac{1-\eqphi{s}}{1+\alpha_f}.
\end{equation}
Therefore the leading order approximation of the coefficients $\zeta_i$ in~(\ref{lambdaeq}) is:
\begin{subequations}
	\begin{align}
	\pert{\zeta_1}{0}=\frac{(\pert{\eqphi{n}}0)^{-1}}{1+\alpha_f}\left[1+\myfrac[2pt]{\alpha_f(\pert{\eqphi{n}}0)^{\theta+1}}{\D \eqphi{s}}\right],\\
	\zeta^{(0)}_{20} = \frac{(\pert{\eqphi{n}}0)^{-1}}{1+\alpha_f}\left(\pert{B_{\phi_s}}0 \eqphi{s} (1-\eqphi{s})+\alpha_f\tilde{\phi}_+\pert{A_{\phi_+}}{-2}\left[\D+\alpha_f(\pert{\eqphi{n}}0)\eqphi{s}^{-1}\right]\right),\\
	\zeta^{(0)}_{22} =\eqphi{s} \pert{\eqphi{n}}0(1+\alpha_f),\\[3pt]
	\zeta^{(0)}_{30}=\alpha_f\eqphi{s}\tilde{\phi}_+\pert{A_{\phi_+}}{-2}\pert{B_{\phi_s}}0, \\
	  \zeta^{(0)}_{32}= \alpha_f\eqphi{s}\tilde{\phi}_+\pert{A_{\phi_+}}{-2}\left(1-\eqphi{s}\right).
	\end{align}
\end{subequations} 
Note here that the above holds for $\alpha_f\gg \phi_0$. When considering instead the case $\alpha_f=O(\phi_0)$, we now have $\epsilon\sim1$ with $\alpha_f\rightarrow0$, the coefficients $\pert{B_{\phi_s}}0$, $\pert{B_{\phi_+}}0$ and $\pert{A_{\phi_s}}0$ can be obtained simply setting $\alpha_f=0$. For what concern $A_{\phi_+}$ we must be more careful to account for all the contribution proportional to $1/\alpha_f$:
\begin{equation} 
A_{\phi_+}=\frac{1}{\tilde{\phi}_+ \alpha_f}\myfrac[1pt]{2\tilde{\phi}_++1-\bar{\phi}_s}{\tilde{\phi}_++1-\bar{\phi}_s}+O(1).
\end{equation}
As mentioned in the main text, this sub-limit does not affect the boundary of the stability region as the dominant term in $A_{\phi_+}$ is always positive.

\section*{Acknowledgments}
MH recognizes support from the Mathematical Institute through a Hooke fellowship, and GC acknowledges the EPSRC and MRC Centre for Doctoral Training in Systems Approaches to Biomedical Science and Cancer Research UK for funding.

\end{document}